\documentclass[aps,amsmath,amssymb,prx]{revtex4-2}  
\usepackage{graphicx}  % needed for figures
\usepackage{dcolumn}   % needed for some tables
\usepackage{bm}        % for math
\usepackage{amssymb}   % for math
\usepackage{lineno}

\usepackage{hyperref}
\newcommand\diff{\,\mathrm{d}}

\def\sign{\sigma_{\chi,n}}

\def\sige{\sigma_{\chi,e}}

\def\sigv{\langle\sigma_{\chi,\chi} v\rangle}

\def\nuflub8{\phi^\nu_B}
\def\nuflube7{\phi^\nu_{Be}}

\def\rhodm{\rho_{\chi}}
\def\rdm{r_{\chi}}
\def\mdm{m_{\chi}}

\def\Ldm{L_{\chi}}
\def\udm{u_{\chi}}
\def\lesim{\lesssim}

\def\vstar{v_{*}}
\def\Vstar{V_{*}}

\def\vesc{v_{\mathrm{esc}}}

\def\Rstar{R_{*}}
\def\Lstar{L_{*}}
\def\rhostar{\rho_{*}}
\def\Mstar{M_{*}}

\def\ndm{n_{\chi}}
\def\Ndm{N_{\chi}}
\def\Ldm{L_{\chi}}

\def\Pobs{P_{\mathrm{obs}}}
\def\dPobs{\dot{P}_{\mathrm{obs}}}

\def\Pthe{P_{\mathrm{the}}}

\def\dPthe{\dot{P}_{\mathrm{the}}}
\def\Msun{M_{\odot}}
\def\Rsun{R_{\odot}}
\def\Lsun{L_{\odot}}
\def\vs{v_{s}}
\def\vt{v_{t}}
\def\dP{\dot{P}}
\def\Ein{E^\mathrm{in}}

\def\Eout{E^\mathrm{out}}

\def\Enet{E^\mathrm{net}}
\def\nat{Nature\ }
\def\aap{Astron.\ Astrophys.\ }
\def\aapr{Astron.\ Astrophys.\ Rev.\ }
\def\apj{Astrophys.\ J.\ }
\def\apjl{Astrophys.\ J.\ Lett.\ }

\def\aj{Astron.\ J.\ }
\def\mnras{Mon.\ Not.\ Roy.\ Astron.\ Soc.\ }

\def\prd{Phys.\ Rev.\ D\ }
\def\prl{Phys.\ Rev.\ Lett.\ }

\def\araa{Annu.\ Rev.\ Astron.\ Astrophys.\ }
\def\jcap{J.\ Cosmol.\ Astropart.\ Phys.\ }

\def\pasp{Publications\ of\ the\ Astronomical\ Society\ of\ the\ Pacific}
\def\epjc{European\ Physical\ Journal\ C}
\def\na{New\ Astronomy\ }

\def\kpc{\,\mathrm{kpc}}
\def\km{\,\mathrm{km}}
\def\TeV{\,\mathrm{TeV}}
\def\GeV{\,\mathrm{GeV}}
\def\MeV{\,\mathrm{MeV}}

\def\MV\TeV{\,\mathrm{MV}}
\def\cm{\,\mathrm{cm}}
\def\s{\,\mathrm{s}}

\newcolumntype{p}{D{,}{\pm}{-1}}

\def\erg{\,\mathrm{erg}}

\begin{document}

% The following information is for internal review, please remove them for submission
%\widetext
%\leftline{Version xx as of \today}
%\leftline{Primary authors: Joe E. Physics}
%\leftline{To be submitted to (PRL, PRD-RC, PRD, PLB; choose one.)}
%\leftline{Comment to {\tt d0-run2eb-nnn@fnal.gov} by xxx, yyy}
%\centerline{\em D\O\ INTERNAL DOCUMENT -- NOT FOR PUBLIC DISTRIBUTION}

% the following line is for submission, including submission to the arXiv!!
%\hspace{5.2in} \mbox{Fermilab-Pub-04/xxx-E}

\title{Possible Dark Matter Signals from White Dwarfs}

\author{Jia-Shu Niu}
\email{jsniu@sxu.edu.cn}
\affiliation{Institute of Theoretical Physics, Shanxi University, Taiyuan, 030006, China}
\affiliation{State Key Laboratory of Quantum Optics and Quantum Optics Devices, Shanxi University, Taiyuan 030006, China}
\affiliation{Collaborative Innovation Center of Extreme Optics, Shanxi University, Taiyuan 030006, China}
\author{Hui-Fang Xue}
\affiliation{Department of Physics, Taiyuan Normal University, Jinzhong, 030619, China}
\affiliation{Institute of Computational and Applied Physics, Taiyuan Normal University, Jinzhong 030619, China}

\date{\today}

\begin{abstract}
In our galaxy, the white dwarfs (WDs) will inevitably capture the dark matter (DM) particles streaming through them, if there exist interactions between DM particles and nuclei/electrons. At the same time, these DM particles can also be evaporated by the nuclei/electrons in a WD if they have proper mass and the WD is not too cold. The evaporation of DM particles will lead to a faster cooling evolution than that predicted by the stellar evolution theory.
In this work, we ascribe the faster cooling evolution of three observed WDs to the capture and evaporation of DM particles, and get the possible DM particle's mass and DM-electron cross section as follows: for $F(q) = 1$, $40 \MeV/c^{2} \lesim \mdm \lesim 70 \MeV/c^{2}$ and $10^{-57} \cm^{2} \lesim \sige \lesim 10^{-55} \cm^{2}$; for $F(q) = (\alpha m_{e})^{2}/q^{2}$, $30 \MeV/c^{2} \lesim \mdm \lesim 60 \MeV/c^{2}$ and $10^{-53} \cm^{2} \lesim \sige \lesim 10^{-51} \cm^{2}$. 
These results are beyond the detection capabilities of current direct detection experiments and should be cross checked by more novel scenarios in the future.
\end{abstract}

%\pacs{}
\maketitle

\section{Motivation}
Although dark matter (DM) is the dominating part of the matter contents of the Universe, the particle nature of DM remains largely unknown. In recent years, the DM particle candidates are searched via three main strategies, i.e., direct detection, indirect detection, and collider searches (see, e.g., Refs. \cite{Nature_dd,Nature_id,Nature_collider} for reviews). Although there are some hints in these detections (see, e.g., Refs. \cite{DAMA01,DAMA02,Cuoco2017,Cui2017,Niu2018b,Niu2019a} and the references therein), null confirmed evidences have been obtained yet.
Currently, the most stringent constraints on elastic scattering DM-nucleon cross section is $\sign \lesim 9.2 \times 10^{-48} \cm^{2}$ for the DM particle's mass $\mdm = 36 \GeV/c^{2}$ (spin-independent), and that for DM-electron is $\sige < 2.1 \times 10^{-41} \cm^{2}$ for $\mdm = 200 \MeV/c^{2}$ ($F(q)=1$) \cite{LZ2023,PandaX-4T2023}.

As the final evolutionary state of most stars in our galaxy \cite{Fontaine2001}, white dwarfs (WDs) have relatively simple interior structures (which is composed by an electron-degenerate core and an atmosphere envelope) and are thought to be the most promising laboratories to measure the DM-electron interactions \cite{Isern2022}.

In our galaxy, the DM particles are inevitably streaming through these WDs, which will loose energy when they scatter with the constituents (nuclei and electrons) of the WDs. If the velocity of these DM particles (after speeding down) less than the escape velocity of a WD, they will be captured and bounded by the star.
These captured DM particles will then accumulate, annihilate in the WD, or evaporate from the WD. All these processes will disturb the normal evolutionary process described by the standard stellar evolution theory.

For the pulsating WDs, the interior structures can be precisely determined according to the asteroseismology on them \cite{Winget2008,Fontaine2008,Althaus2010,Calcaferro2017}, which provide us the detailed information to calculate the processes of DM capture, evaporation, and annihilation. More importantly, the evolutionary state can be not only calculated by the stellar evolution theory, but also indicated by the secular period variation rates of these WDs' pulsation modes from observations.
As a result, the effects of the DM particles on a pulsating WD can be represented by the differences of the period variation rate of the star between the theory and observations.

In general, the period variation rate of a pulsating WD's pulsation mode ($\dP \equiv \diff P/\diff t$) can be expressed as \cite{Winget1983}
\begin{equation}
  \label{eq:WD_rate_R}
  \frac{\dP}{P} \simeq -\frac{1}{2} \frac{\dot{T}_{c}}{T_{c}} + \frac{\dot{R}_{*}}{\Rstar},
\end{equation}
where $P$ is the period of the pulsation mode, $T_{c}$ is the core temperature of the WD, and $\Rstar$ is the radius of the WD. If we use the mass-radius relation of the low mass WDs ($\Rstar \propto \Mstar^{-\frac{1}{3}}$), the above expression can be expressed as
\begin{equation}
  \label{eq:WD_rate_M}
  \frac{\dP}{P} \simeq -\frac{1}{2} \frac{\dot{T}_{c}}{T_{c}} - \frac{1}{3}\frac{\dot{M}_{*}}{\Mstar},
\end{equation}
where $\Mstar$ is the mass of the WD.

Based on Eq. (\ref{eq:WD_rate_M}), the effects of DM related processes on a pulsating WD are clear\footnote{Here, we take $P$, $T_{c}$, and $\Mstar$ as constants.}: (a) DM capture and accumulation will decrease $\dP$, in which processes the DM particles transfer kinetic energy to the star's constituents ($\dot{T}_{c} > 0$) and increase the mass in the star ($\dot{M}_{*} >0$); (b) DM evaporation will increase $\dP$, in which process the star's constituents transfer kinetic energy to the DM particles ($\dot{T}_{c} < 0$) and decrease the mass in the star ($\dot{M}_{*} <0$); (c) DM annihilation will decrease $\dP$, in which process the DM particles inject energy into the star ($\dot{T}_{c} > 0$) (see, e.g. Ref. \cite{Niu2018_dav}).

What is interesting comes from the observation of the hydrogen-atmosphere pulsating white dwarfs (DAVs or ZZ Cetis, see more details in Refs. \cite{Brickhill1983,Mukadam2006}). Up to now, the period variation rates of some DAVs (G117-B15A, R548, and L19-2) have been already determined by long-term time-series photometric observations. However, almost all the observed period variation rates of the stable pulsation modes of these DAVs are greater than those predicted from stellar evolution theory (see Table \ref{tab:DAVs}). This phenomenon could be explained by the additional cooling mechanism like axion (see, e.g. Refs. \cite{Corsico2016,Romero2012,Sullivan2015}).

In this work, we assume that all these discrepancies between the observation and stellar evolutionary theory are caused by the elastic scattering between the DM particles and the constituents of the DAVs.
Considering the above discussion about the effects of DM particles, a larger period variation rate (corresponding to faster cooling evolution) can only be caused by the efficient evaporation process.

\begin{table*}[htp!]
  \centering
  \caption{Information of the Three DAVs.}
  \label{tab:DAVs}
  % \scalebox{0.8}{
%%\resizebox{1.0\textwidth}{!}{
  \begin{ruledtabular}
  \begin{tabular}{l|ccc}
%%    \hline
%%    \hline
ID    &{G117-B15A}                   & {R548}                   & {L19-2} \\
Marks    &{DAV1}                   & {DAV2}                   & {DAV3} \\
    \hline
    $\Pobs\ (\s)$                  & 215.20 & 212.95 & 113.8\\
    $\Pthe\ (\s)$                  & 215.215 & 213.401 & 113.41\\
    ${\dPobs/\Pobs}$ $(\s/\s)$ & $(5.12\pm0.82)\times10^{-15}$ &$(3.3\pm1.1)\times10^{-15}$ & $(3.0\pm0.6)\times10^{-15}$ \\
    ${\dPthe/\Pthe}$ $(\s/\s)$ & $1.25\times10^{-15}$ &$1.08\times10^{-15}$ & $1.42\times10^{-15}$\\
    $\Mstar/\Msun$                        & $0.593\pm0.007$ & $0.609\pm0.012$ & $0.705\pm0.023$\\
    $\log{(\Lstar/\Lsun)}$         & $-2.497\pm0.030$ & $-2.594\pm0.025$ & $-2.622\pm0.046$ \\
    $\log{(\Rstar/\Rsun)}$         & $-1.882\pm0.029$ & $-1.904\pm0.015$ & $-1.945\pm0.037$\\
    Distance* ($\mathrm{pc}$)   & 57.37& 32.71  & 20.87 \\
\hline
Refs.& \cite{Romero2012,Kepler2021,Bailer2021}& \cite{Romero2012,Mukadam2013,Kepler2021,Bailer2021}& \cite{Sullivan2015,Corsico2016,Pajdosz1995,Bailer2021} \\
%%\hline
  \end{tabular}
\end{ruledtabular}
%%  \end{center}
%%}
\\
\footnotesize{Note: $\Pobs$ is the period of a specific pulsation mode from observation, $\dPobs$ is its variation rate;  $\Pthe$ is the period of a specific pulsation mode from stellar evolution theory, $\dPthe$ is its variation rate; $\Mstar$, $\Msun$ are the mass of the DAV and Sun; $\Lstar$, $\Lsun$ are the luminosity of the DAV and Sun; $\Rstar$, $\Rsun$ are the radius of the DAV and Sun; the distances of the DAVs to the Sun are obtained based on Gaia DR3 \cite{Bailer2021}.}
\end{table*}

\section{Capture and Evaporation of DM Particles in DAVs}
When the galactic DM particles are streaming through a DAV, some of them will loose energy and be captured by the star. At the same time, the captured DM particles will get enough energy and be evaporated from the star.\footnote{Here, we consider the non-annihilating DM particles. The DM pair annihilation effect will be discussed later in this work.}
The evolution of the total number of DM particles inside the star ($\Ndm$) can be written as,
\begin{equation}
  \frac{\diff \Ndm}{\diff t} = C_{*} - E_{*} \cdot \Ndm,
\label{eq:Ndm}
\end{equation}
where $C_{*}$ is the DM particle capture rate of the star, and $E_{*}$ is the DM particle evaporation rate of the star. The solution of Eq. (\ref{eq:Ndm}) is 
\begin{equation}
  \Ndm(t) = C_{*} t \cdot (\frac{1-e^{-E_{*}t}}{E_{*}t}).
\label{eq:sol_cap_eva}
\end{equation}

If the evaporation is the dominating effect, the equilibrium state between capture and evaporation will be researched in a small time scale ($\sim 1/E_{*}$) compared to the time scale of stellar evolution. At the equilibrium state, both the DM particle numbers captured and evaporated in a unit of time by the star are $C_{*}$. These DM particles provide a new energy transfer channel between the star and the external environment, which will change the standard cooling evolution of the DAV.

In a DAV, the capture and evaporation rates ($C_{*}$ and $E_{*}$) for nuclei and electrons are different because of the different interaction cross sections and different equations of state (most of the electrons are in the Fermi degenerate state). 
The detailed calculation of $C_{*}$ and $E_{*}$ for nuclei and electrons can be found in Appendix \ref{app:01}.

In a DAV which is in the equilibrium state between capture and evaporation of DM particles, the capture process will transfer the energy of DM to the star (marked as $\Ein$), while the evaporation process will transfer the energy of star to the DM and then the external environment (marked as $\Eout$). In general, the energy transfer of these two processes are not equal to each other, and a DAV will lose or gain net energy ($\Enet \equiv \Eout - \Ein$) in the equilibrium state via DM particles. The observed results of DAVs demand that $\Enet > 0$, i.e. the DAVs should lose energy in the equilibrium state. 
The detailed expressions of $\Ein$ and $\Eout$ for nuclei and electrons can be found in Appendix \ref{app:02}.

\begin{figure}[htp!]
  \centering
  \includegraphics[width=0.8\textwidth]{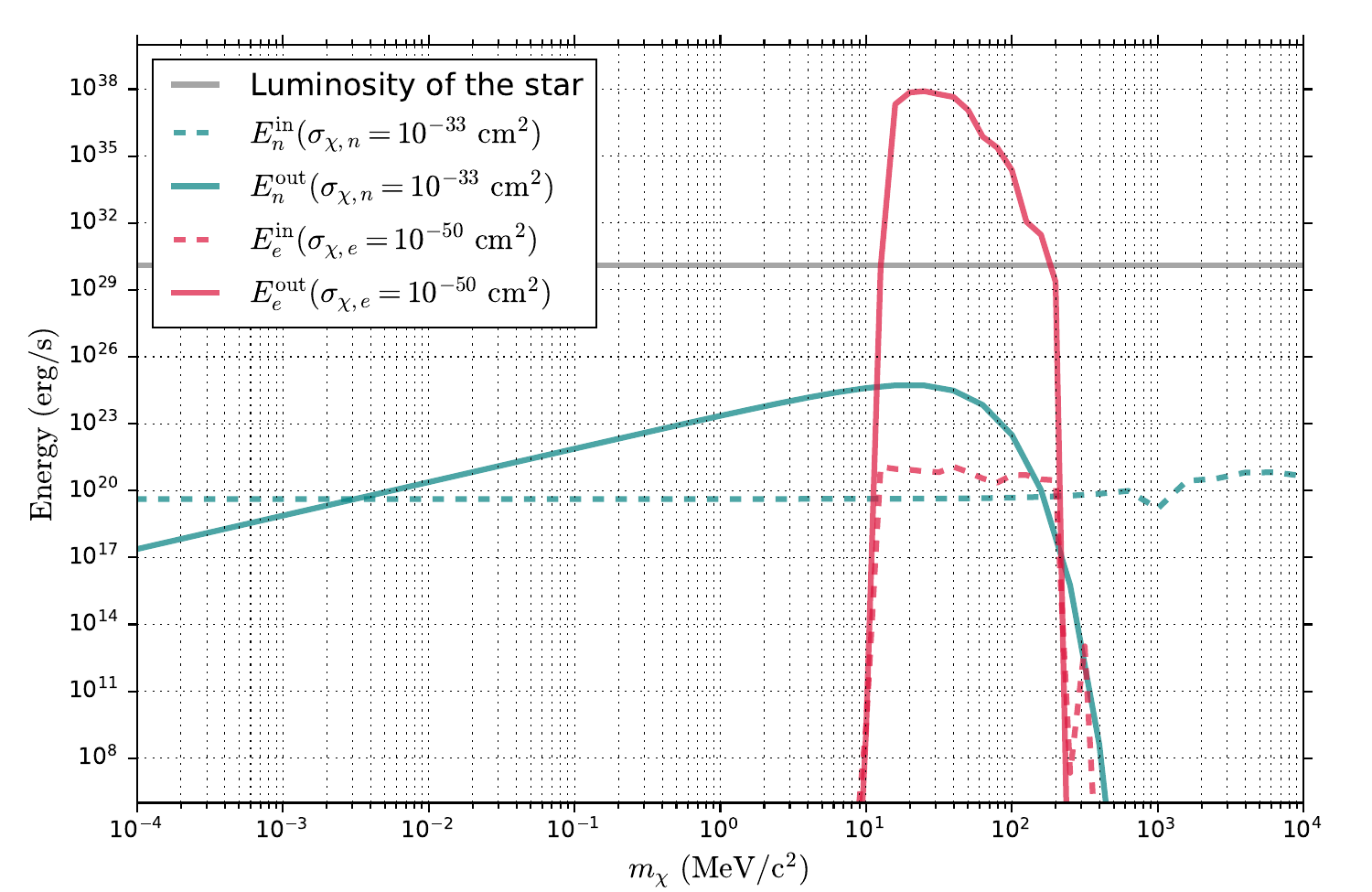}
  \caption{Energy gaining ($\Ein$) and losing ($\Eout$) of nuclei and electrons in DAV1 in a unit of time. The luminosity of the star is also marked, which can be considered as the threshold to reproduce the observed period variation rates. }
  \label{fig:E_in_out}
\end{figure}

Taking DAV1 as an example, Figure \ref{fig:E_in_out} shows the energy gaining and losing of nuclei and electrons in WD1, for the specific cross sections: $\sign = 10^{-33}\ \cm^{2}$ and $\sige = 10^{-50}\ \cm^{2}$ ($F(q) = 1$ for both cases, and the capture and evaporation of nuclei and electrons are calculated independently). It indicates that electrons are more efficient in transferring DM particle energy than nuclei. Considering the current limits on $\sign$ ($\lesim 10^{-33} \cm^2$ for $1-10^{3} \MeV/c^{2}$ \cite{DMMeV2023}), it cannot reproduce the observed results based on the DM evaporation caused by the nuclei, which should transfer energy comparable with the luminosity of the star. Consequently, we will only consider the interaction between DM particles and electrons in the rest of this work.

\section{DM Effects on Period Variation Rates of DAVs}
In the equilibrium state between capture and evaporation, the DM particle number in a DAV is a constant ($C_{*}/E_{*}$), and then $\dot{M}_{*} = 0$. Based on Eq. (\ref{eq:WD_rate_M}), if $\Enet > 0$ (which lead $\dot{T}_{c}<0$), we get a larger $\dot{P}$ which could meet the observed results of the DAVs. In this situation, the transferred net energy can be considered as another form of luminosity which is produced by DM particles ($\Ldm \equiv \Enet$).

Similar to the situation of axions \cite{Isern1992,Corsico2001}, we have 
\begin{equation}
  \label{eq:dPdL}
  \frac{\dPobs}{\dPthe} = \frac{\Lstar+\Ldm}{\Lstar},
\end{equation}
where $\dPthe$ is the period variation rate predicted by the stellar evolution theory; $\dPobs$ is the observed period variation rate; $\Lstar$ is the luminosity from asteroseismology model.

\section{Results and Discussions}
\begin{figure*}[htp!]
  \centering
  \includegraphics[width=0.8\textwidth]{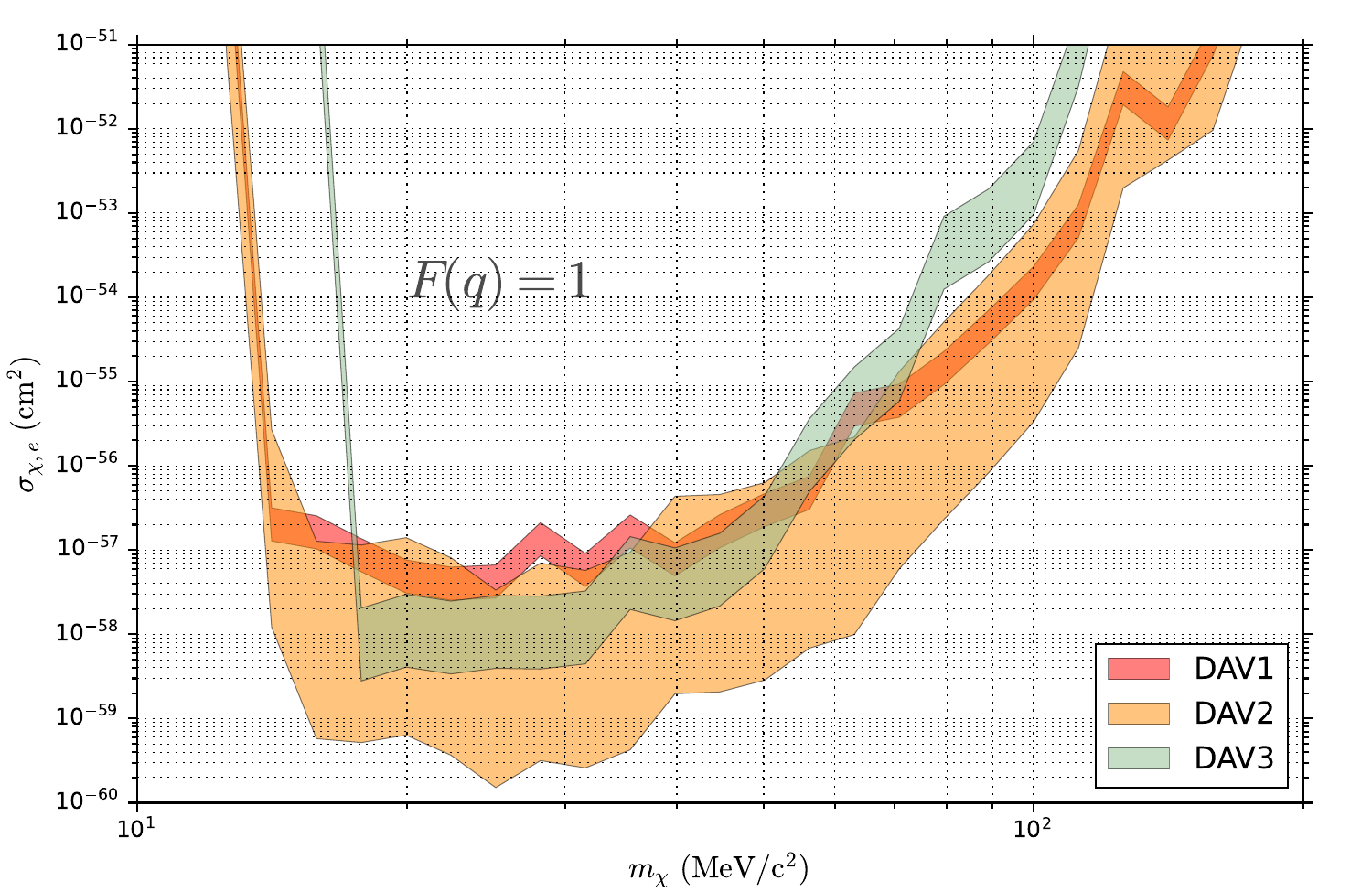}
  \includegraphics[width=0.8\textwidth]{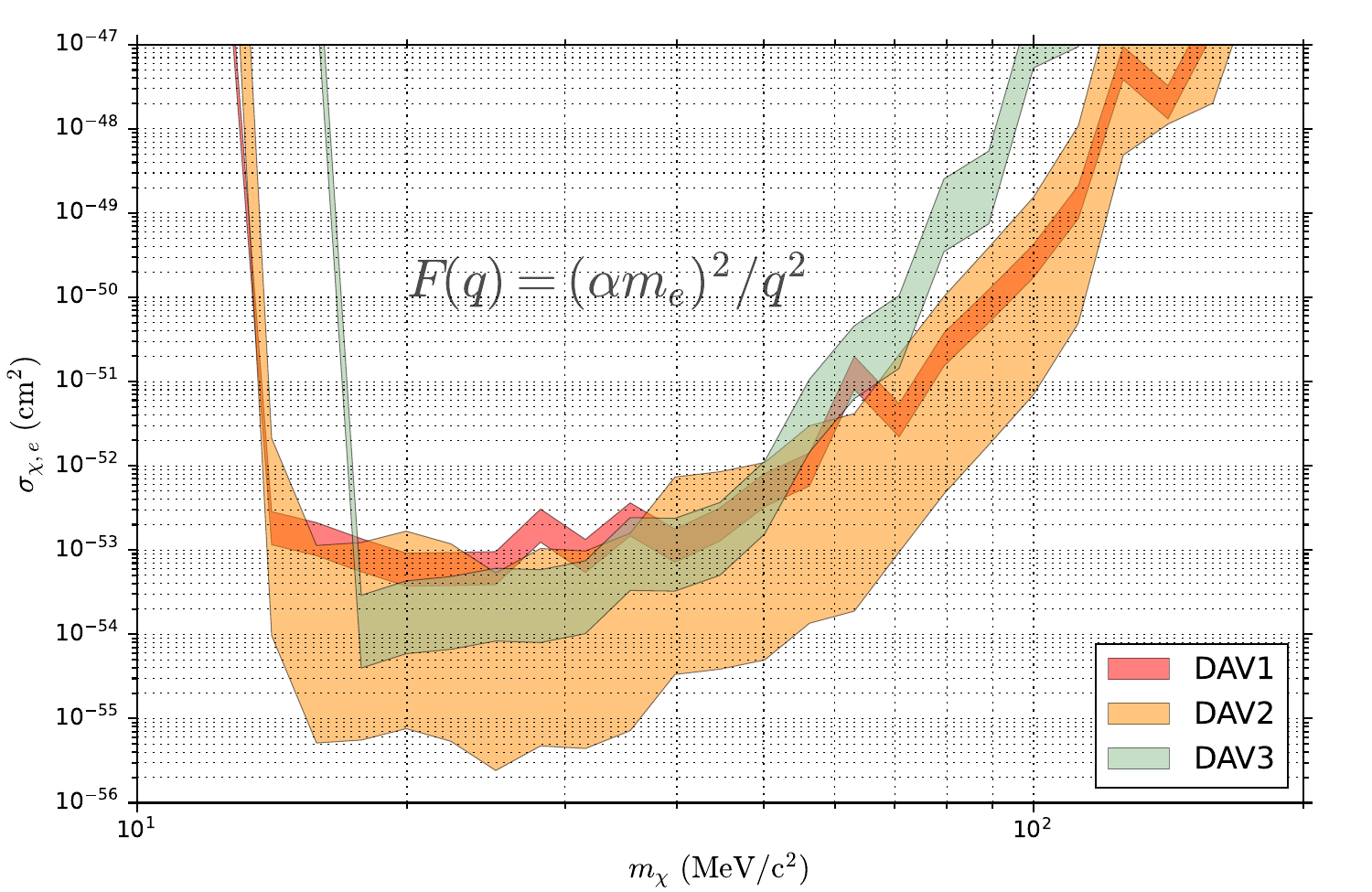}
  \caption{Parameter space of DM particles obtained from the three DAVs. Both $F(q) = 1$ and $F(q) = (\alpha m_{e})^{2}/q^{2}$ cases are considered.}
  \label{fig:sigma_mass}
\end{figure*}

Based on the observed period variation rates of the three DAVs, we can use Eq. (\ref{eq:dPdL}) to obtain the DM parameter space of them independently. In Figure \ref{fig:sigma_mass}, the allowed $\pm 2\sigma$ DM parameter spaces are presented in two DM form factor cases ($F(q)=1$ and $F(q) = (\alpha m_{e})^{2}/q^{2}$), and each of them is a banded area. 

The intersecting regions in Figure \ref{fig:sigma_mass} provide us the possible DM mass and DM-electron cross section ranges: 
\begin{itemize}
  \item {for $F(q) = 1$, $40 \MeV/c^{2} \lesim \mdm \lesim 70 \MeV/c^{2}$ and $10^{-57} \cm^{2} \lesim \sige \lesim 10^{-55} \cm^{2}$;}
  \item {for $F(q) = (\alpha m_{e})^{2}/q^{2}$, $30 \MeV/c^{2} \lesim \mdm \lesim 60 \MeV/c^{2}$ and $10^{-53} \cm^{2} \lesim \sige \lesim 10^{-51} \cm^{2}$.}
\end{itemize}

The energy injection in the star according to annihilation process can be estimated by 
\begin{equation}
  \label{eq:E_ann}
  \Ein_{ann} = \int_{0}^{\Rstar} 4 \pi r^{2} \sigv (n_{0} \cdot \ndm(r))^{2} \diff r,
\end{equation}
where $\sigv \simeq 3\times 10^{-26} \cm^{3} \s^{-1}$ is the velocity-averaged DM annihilation cross section ($\sigma_{\chi,\chi}$) multiplied by DM relative velocity ($v$); $n_{0} = C_{*}/E_{*}$ is the DM particles in the capture-evaporation equilibrium state; $\ndm(r)$ is the DM particle distribution in the star. When $\mdm = 40 \MeV$, $\Ein_{ann} \simeq 10^{-7} \erg/\s$ for the nucleus case ($\sign = 10^{-33} \cm^{2}$) and $\Ein_{ann} \simeq 10^{-31} \erg/\s$ for the electron case ($\sige = 10^{-50} \cm^{2}$), which is much less than the transferred energy by capture and evaporation processes (see in Figure \ref{fig:E_in_out}). Consequently, the results of this work also hold for the DM particles that can annihilate by pairs.

Benefited from the Fermi degenerate state of the electrons in WDs and the large escape velocity, we can explore the DM parameter spaces which cannot be reached by current DM direct detection experiments. The novel scenario in this work has not only the potential to be developed more precisely (if we consider the structure details of the star and obtain more precise $\dPobs$ by high-precision time-series photometric observations), but also the potential to be extended to much more DAVs, and even other type of pulsating stars who have faster evolution states than the stellar evolution theory predicted (see, e.g., Refs. \cite{Niu2022,Xue2022}).
We hope that more novel scenarios could be proposed and implemented to perform cross checks on the results of this work.

%%\section{Acknowledgments}
\begin{acknowledgments}
  J.S.N. acknowledges support from the National Natural Science Foundation of China (NSFC) (No. 12005124 and No. 12147215). H.F.X. acknowledges support from the National Natural Science Foundation of China (NSFC) (No. 12303036) and the Applied Basic Research Programs of Natural Science Foundation of Shanxi Province (No. 202103021223320).
\end{acknowledgments}

%%\bibliographystyle{prl}
%%\bibliography{dm_astero}% Produces the bibliography via BibTeX.

\begin{thebibliography}{39}%
\makeatletter
\providecommand \@ifxundefined [1]{%
 \@ifx{#1\undefined}
}%
\providecommand \@ifnum [1]{%
 \ifnum #1\expandafter \@firstoftwo
 \else \expandafter \@secondoftwo
 \fi
}%
\providecommand \@ifx [1]{%
 \ifx #1\expandafter \@firstoftwo
 \else \expandafter \@secondoftwo
 \fi
}%
\providecommand \natexlab [1]{#1}%
\providecommand \enquote  [1]{``#1''}%
\providecommand \bibnamefont  [1]{#1}%
\providecommand \bibfnamefont [1]{#1}%
\providecommand \citenamefont [1]{#1}%
\providecommand \href@noop [0]{\@secondoftwo}%
\providecommand \href [0]{\begingroup \@sanitize@url \@href}%
\providecommand \@href[1]{\@@startlink{#1}\@@href}%
\providecommand \@@href[1]{\endgroup#1\@@endlink}%
\providecommand \@sanitize@url [0]{\catcode `\\12\catcode `\$12\catcode
  `\&12\catcode `\#12\catcode `\^12\catcode `\_12\catcode `\%12\relax}%
\providecommand \@@startlink[1]{}%
\providecommand \@@endlink[0]{}%
\providecommand \url  [0]{\begingroup\@sanitize@url \@url }%
\providecommand \@url [1]{\endgroup\@href {#1}{\urlprefix }}%
\providecommand \urlprefix  [0]{URL }%
\providecommand \Eprint [0]{\href }%
\providecommand \doibase [0]{https://doi.org/}%
\providecommand \selectlanguage [0]{\@gobble}%
\providecommand \bibinfo  [0]{\@secondoftwo}%
\providecommand \bibfield  [0]{\@secondoftwo}%
\providecommand \translation [1]{[#1]}%
\providecommand \BibitemOpen [0]{}%
\providecommand \bibitemStop [0]{}%
\providecommand \bibitemNoStop [0]{.\EOS\space}%
\providecommand \EOS [0]{\spacefactor3000\relax}%
\providecommand \BibitemShut  [1]{\csname bibitem#1\endcsname}%
\let\auto@bib@innerbib\@empty
%</preamble>
\bibitem [{\citenamefont {{Liu}}\ \emph {et~al.}(2016)\citenamefont {{Liu}},
  \citenamefont {{Chen}},\ and\ \citenamefont {{Ji}}}]{Nature_dd}%
  \BibitemOpen
  \bibfield  {author} {\bibinfo {author} {\bibfnamefont {J.}~\bibnamefont
  {{Liu}}}, \bibinfo {author} {\bibfnamefont {X.}~\bibnamefont {{Chen}}},\ and\
  \bibinfo {author} {\bibfnamefont {X.}~\bibnamefont {{Ji}}},\ }\bibfield
  {title} {\bibinfo {title} {{Current status of direct dark matter detection
  experiments}},\ }\href {https://doi.org/10.1038/nphys4039} {\bibfield
  {journal} {\bibinfo  {journal} {Nature Physics}\ }\textbf {\bibinfo {volume}
  {13}},\ \bibinfo {pages} {212} (\bibinfo {year} {2016})}\BibitemShut
  {NoStop}%
\bibitem [{\citenamefont {{Conrad}}\ and\ \citenamefont
  {{Reimer}}(2017)}]{Nature_id}%
  \BibitemOpen
  \bibfield  {author} {\bibinfo {author} {\bibfnamefont {J.}~\bibnamefont
  {{Conrad}}}\ and\ \bibinfo {author} {\bibfnamefont {O.}~\bibnamefont
  {{Reimer}}},\ }\bibfield  {title} {\bibinfo {title} {{Indirect dark matter
  searches in gamma and cosmic rays}},\ }\href
  {https://doi.org/10.1038/nphys4049} {\bibfield  {journal} {\bibinfo
  {journal} {Nature Physics}\ }\textbf {\bibinfo {volume} {13}},\ \bibinfo
  {pages} {224} (\bibinfo {year} {2017})}\BibitemShut {NoStop}%
\bibitem [{\citenamefont {{Buchmueller}}\ \emph {et~al.}(2017)\citenamefont
  {{Buchmueller}}, \citenamefont {{Doglioni}},\ and\ \citenamefont
  {{Wang}}}]{Nature_collider}%
  \BibitemOpen
  \bibfield  {author} {\bibinfo {author} {\bibfnamefont {O.}~\bibnamefont
  {{Buchmueller}}}, \bibinfo {author} {\bibfnamefont {C.}~\bibnamefont
  {{Doglioni}}},\ and\ \bibinfo {author} {\bibfnamefont {L.-T.}\ \bibnamefont
  {{Wang}}},\ }\bibfield  {title} {\bibinfo {title} {{Search for dark matter at
  colliders}},\ }\href {https://doi.org/10.1038/nphys4054} {\bibfield
  {journal} {\bibinfo  {journal} {Nature Physics}\ }\textbf {\bibinfo {volume}
  {13}},\ \bibinfo {pages} {217} (\bibinfo {year} {2017})}\BibitemShut
  {NoStop}%
\bibitem [{\citenamefont {{Bernabei}}\ \emph {et~al.}(2008)\citenamefont
  {{Bernabei}}, \citenamefont {{Belli}}, \citenamefont {{Cappella}},
  \citenamefont {{Cerulli}}, \citenamefont {{Dai}}, \citenamefont {{D'Angelo}},
  \citenamefont {{He}}, \citenamefont {{Incicchitti}}, \citenamefont {{Kuang}},
  \citenamefont {{Ma}}, \citenamefont {{Montecchia}}, \citenamefont
  {{Nozzoli}}, \citenamefont {{Prosperi}}, \citenamefont {{Sheng}},\ and\
  \citenamefont {{Ye}}}]{DAMA01}%
  \BibitemOpen
  \bibfield  {author} {\bibinfo {author} {\bibfnamefont {R.}~\bibnamefont
  {{Bernabei}}}, \bibinfo {author} {\bibfnamefont {P.}~\bibnamefont {{Belli}}},
  \bibinfo {author} {\bibfnamefont {F.}~\bibnamefont {{Cappella}}}, \bibinfo
  {author} {\bibfnamefont {R.}~\bibnamefont {{Cerulli}}}, \bibinfo {author}
  {\bibfnamefont {C.~J.}\ \bibnamefont {{Dai}}}, \bibinfo {author}
  {\bibfnamefont {A.}~\bibnamefont {{D'Angelo}}}, \bibinfo {author}
  {\bibfnamefont {H.~L.}\ \bibnamefont {{He}}}, \bibinfo {author}
  {\bibfnamefont {A.}~\bibnamefont {{Incicchitti}}}, \bibinfo {author}
  {\bibfnamefont {H.~H.}\ \bibnamefont {{Kuang}}}, \bibinfo {author}
  {\bibfnamefont {J.~M.}\ \bibnamefont {{Ma}}}, \bibinfo {author}
  {\bibfnamefont {F.}~\bibnamefont {{Montecchia}}}, \bibinfo {author}
  {\bibfnamefont {F.}~\bibnamefont {{Nozzoli}}}, \bibinfo {author}
  {\bibfnamefont {D.}~\bibnamefont {{Prosperi}}}, \bibinfo {author}
  {\bibfnamefont {X.~D.}\ \bibnamefont {{Sheng}}},\ and\ \bibinfo {author}
  {\bibfnamefont {Z.~P.}\ \bibnamefont {{Ye}}},\ }\bibfield  {title} {\bibinfo
  {title} {{First results from DAMA/LIBRA and the combined results with
  DAMA/NaI}},\ }\href {https://doi.org/10.1140/epjc/s10052-008-0662-y}
  {\bibfield  {journal} {\bibinfo  {journal} {\epjc}\ }\textbf {\bibinfo
  {volume} {56}},\ \bibinfo {pages} {333} (\bibinfo {year} {2008})},\ \Eprint
  {https://arxiv.org/abs/0804.2741} {arXiv:0804.2741 [astro-ph]} \BibitemShut
  {NoStop}%
\bibitem [{\citenamefont {{Bernabei}}\ \emph {et~al.}(2010)\citenamefont
  {{Bernabei}}, \citenamefont {{Belli}}, \citenamefont {{Cappella}},
  \citenamefont {{Cerulli}}, \citenamefont {{Dai}}, \citenamefont {{D'Angelo}},
  \citenamefont {{He}}, \citenamefont {{Incicchitti}}, \citenamefont {{Kuang}},
  \citenamefont {{Ma}}, \citenamefont {{Montecchia}}, \citenamefont
  {{Nozzoli}}, \citenamefont {{Prosperi}}, \citenamefont {{Sheng}},
  \citenamefont {{Wang}},\ and\ \citenamefont {{Ye}}}]{DAMA02}%
  \BibitemOpen
  \bibfield  {author} {\bibinfo {author} {\bibfnamefont {R.}~\bibnamefont
  {{Bernabei}}}, \bibinfo {author} {\bibfnamefont {P.}~\bibnamefont {{Belli}}},
  \bibinfo {author} {\bibfnamefont {F.}~\bibnamefont {{Cappella}}}, \bibinfo
  {author} {\bibfnamefont {R.}~\bibnamefont {{Cerulli}}}, \bibinfo {author}
  {\bibfnamefont {C.~J.}\ \bibnamefont {{Dai}}}, \bibinfo {author}
  {\bibfnamefont {A.}~\bibnamefont {{D'Angelo}}}, \bibinfo {author}
  {\bibfnamefont {H.~L.}\ \bibnamefont {{He}}}, \bibinfo {author}
  {\bibfnamefont {A.}~\bibnamefont {{Incicchitti}}}, \bibinfo {author}
  {\bibfnamefont {H.~H.}\ \bibnamefont {{Kuang}}}, \bibinfo {author}
  {\bibfnamefont {X.~H.}\ \bibnamefont {{Ma}}}, \bibinfo {author}
  {\bibfnamefont {F.}~\bibnamefont {{Montecchia}}}, \bibinfo {author}
  {\bibfnamefont {F.}~\bibnamefont {{Nozzoli}}}, \bibinfo {author}
  {\bibfnamefont {D.}~\bibnamefont {{Prosperi}}}, \bibinfo {author}
  {\bibfnamefont {X.~D.}\ \bibnamefont {{Sheng}}}, \bibinfo {author}
  {\bibfnamefont {R.~G.}\ \bibnamefont {{Wang}}},\ and\ \bibinfo {author}
  {\bibfnamefont {Z.~P.}\ \bibnamefont {{Ye}}},\ }\bibfield  {title} {\bibinfo
  {title} {{New results from DAMA/LIBRA}},\ }\href
  {https://doi.org/10.1140/epjc/s10052-010-1303-9} {\bibfield  {journal}
  {\bibinfo  {journal} {\epjc}\ }\textbf {\bibinfo {volume} {67}},\ \bibinfo
  {pages} {39} (\bibinfo {year} {2010})},\ \Eprint
  {https://arxiv.org/abs/1002.1028} {arXiv:1002.1028 [astro-ph.GA]}
  \BibitemShut {NoStop}%
\bibitem [{\citenamefont {Cuoco}\ \emph {et~al.}(2017)\citenamefont {Cuoco},
  \citenamefont {Kr\"amer},\ and\ \citenamefont {Korsmeier}}]{Cuoco2017}%
  \BibitemOpen
  \bibfield  {author} {\bibinfo {author} {\bibfnamefont {A.}~\bibnamefont
  {Cuoco}}, \bibinfo {author} {\bibfnamefont {M.}~\bibnamefont {Kr\"amer}},\
  and\ \bibinfo {author} {\bibfnamefont {M.}~\bibnamefont {Korsmeier}},\
  }\bibfield  {title} {\bibinfo {title} {Novel dark matter constraints from
  antiprotons in light of ams-02},\ }\href
  {https://doi.org/10.1103/PhysRevLett.118.191102} {\bibfield  {journal}
  {\bibinfo  {journal} {\prl}\ }\textbf {\bibinfo {volume} {118}},\ \bibinfo
  {pages} {191102} (\bibinfo {year} {2017})}\BibitemShut {NoStop}%
\bibitem [{\citenamefont {Cui}\ \emph {et~al.}(2017)\citenamefont {Cui},
  \citenamefont {Yuan}, \citenamefont {Tsai},\ and\ \citenamefont
  {Fan}}]{Cui2017}%
  \BibitemOpen
  \bibfield  {author} {\bibinfo {author} {\bibfnamefont {M.-Y.}\ \bibnamefont
  {Cui}}, \bibinfo {author} {\bibfnamefont {Q.}~\bibnamefont {Yuan}}, \bibinfo
  {author} {\bibfnamefont {Y.-L.~S.}\ \bibnamefont {Tsai}},\ and\ \bibinfo
  {author} {\bibfnamefont {Y.-Z.}\ \bibnamefont {Fan}},\ }\bibfield  {title}
  {\bibinfo {title} {Possible dark matter annihilation signal in the ams-02
  antiproton data},\ }\href {https://doi.org/10.1103/PhysRevLett.118.191101}
  {\bibfield  {journal} {\bibinfo  {journal} {\prl}\ }\textbf {\bibinfo
  {volume} {118}},\ \bibinfo {pages} {191101} (\bibinfo {year}
  {2017})}\BibitemShut {NoStop}%
\bibitem [{\citenamefont {{Niu}}\ \emph
  {et~al.}(2018{\natexlab{a}})\citenamefont {{Niu}}, \citenamefont {{Li}},
  \citenamefont {{Ding}}, \citenamefont {{Zhu}}, \citenamefont {{Xue}},\ and\
  \citenamefont {{Wang}}}]{Niu2018b}%
  \BibitemOpen
  \bibfield  {author} {\bibinfo {author} {\bibfnamefont {J.-S.}\ \bibnamefont
  {{Niu}}}, \bibinfo {author} {\bibfnamefont {T.}~\bibnamefont {{Li}}},
  \bibinfo {author} {\bibfnamefont {R.}~\bibnamefont {{Ding}}}, \bibinfo
  {author} {\bibfnamefont {B.}~\bibnamefont {{Zhu}}}, \bibinfo {author}
  {\bibfnamefont {H.-F.}\ \bibnamefont {{Xue}}},\ and\ \bibinfo {author}
  {\bibfnamefont {Y.}~\bibnamefont {{Wang}}},\ }\bibfield  {title} {\bibinfo
  {title} {{Bayesian analysis of the break in DAMPE lepton spectra}},\ }\href
  {https://doi.org/10.1103/PhysRevD.97.083012} {\bibfield  {journal} {\bibinfo
  {journal} {\prd}\ }\textbf {\bibinfo {volume} {97}},\ \bibinfo {eid} {083012}
  (\bibinfo {year} {2018}{\natexlab{a}})},\ \Eprint
  {https://arxiv.org/abs/1712.00372} {arXiv:1712.00372 [astro-ph.HE]}
  \BibitemShut {NoStop}%
\bibitem [{\citenamefont {{Niu}}\ \emph {et~al.}(2019)\citenamefont {{Niu}},
  \citenamefont {{Li}},\ and\ \citenamefont {{Xu}}}]{Niu2019a}%
  \BibitemOpen
  \bibfield  {author} {\bibinfo {author} {\bibfnamefont {J.-S.}\ \bibnamefont
  {{Niu}}}, \bibinfo {author} {\bibfnamefont {T.}~\bibnamefont {{Li}}},\ and\
  \bibinfo {author} {\bibfnamefont {F.-Z.}\ \bibnamefont {{Xu}}},\ }\bibfield
  {title} {\bibinfo {title} {{A simple and natural interpretations of the DAMPE
  cosmic-ray electron/positron spectrum within two sigma deviations}},\ }\href
  {https://doi.org/10.1140/epjc/s10052-019-6625-7} {\bibfield  {journal}
  {\bibinfo  {journal} {European Physical Journal C}\ }\textbf {\bibinfo
  {volume} {79}},\ \bibinfo {eid} {125} (\bibinfo {year} {2019})},\ \Eprint
  {https://arxiv.org/abs/1712.09586} {arXiv:1712.09586 [hep-ph]} \BibitemShut
  {NoStop}%
\bibitem [{\citenamefont {{Aalbers}}\ \emph {et~al.}(2023)\citenamefont
  {{Aalbers}}, \citenamefont {{Akerib}}, \citenamefont {{Akerlof}},
  \citenamefont {{Al Musalhi}}, \citenamefont {{Alder}}, \citenamefont
  {{Alqahtani}}, \citenamefont {{Alsum}}, \citenamefont {{Amarasinghe}},
  \citenamefont {{Ames}}, \citenamefont {{Anderson}}, \citenamefont
  {{Angelides}}, \citenamefont {{Ara{\'u}jo}}, \citenamefont {{Armstrong}},
  \citenamefont {{Arthurs}}, \citenamefont {{Azadi}}, \citenamefont {{Bailey}},
  \citenamefont {{Baker}}, \citenamefont {{Balajthy}}, \citenamefont
  {{Balashov}}, \citenamefont {{Bang}}, \citenamefont {{Bargemann}},
  \citenamefont {{Barry}}, \citenamefont {{Barthel}}, \citenamefont {{Bauer}},
  \citenamefont {{Baxter}}, \citenamefont {{Beattie}}, \citenamefont {{Belle}},
  \citenamefont {{Beltrame}}, \citenamefont {{Bensinger}}, \citenamefont
  {{Benson}}, \citenamefont {{Bernard}}, \citenamefont {{Bhatti}},
  \citenamefont {{Biekert}}, \citenamefont {{Biesiadzinski}}, \citenamefont
  {{Birch}}, \citenamefont {{Birrittella}}, \citenamefont {{Blockinger}},
  \citenamefont {{Boast}}, \citenamefont {{Boxer}}, \citenamefont {{Bramante}},
  \citenamefont {{Brew}}, \citenamefont {{Br{\'a}s}}, \citenamefont
  {{Buckley}}, \citenamefont {{Bugaev}}, \citenamefont {{Burdin}},
  \citenamefont {{Busenitz}}, \citenamefont {{Buuck}}, \citenamefont
  {{Cabrita}}, \citenamefont {{Carels}}, \citenamefont {{Carlsmith}},
  \citenamefont {{Carlson}}, \citenamefont {{Carmona-Benitez}}, \citenamefont
  {{Cascella}}, \citenamefont {{Chan}}, \citenamefont {{Chawla}}, \citenamefont
  {{Chen}}, \citenamefont {{Cherwinka}}, \citenamefont {{Chott}}, \citenamefont
  {{Cole}}, \citenamefont {{Coleman}}, \citenamefont {{Converse}},
  \citenamefont {{Cottle}}, \citenamefont {{Cox}}, \citenamefont {{Craddock}},
  \citenamefont {{Creaner}}, \citenamefont {{Curran}}, \citenamefont
  {{Currie}}, \citenamefont {{Cutter}}, \citenamefont {{Dahl}}, \citenamefont
  {{David}}, \citenamefont {{Davis}}, \citenamefont {{Davison}}, \citenamefont
  {{Delgaudio}}, \citenamefont {{Dey}}, \citenamefont {{de Viveiros}},
  \citenamefont {{Dobi}}, \citenamefont {{Dobson}}, \citenamefont
  {{Druszkiewicz}}, \citenamefont {{Dushkin}}, \citenamefont {{Edberg}},
  \citenamefont {{Edwards}}, \citenamefont {{Elnimr}}, \citenamefont {{Emmet}},
  \citenamefont {{Eriksen}}, \citenamefont {{Faham}}, \citenamefont {{Fan}},
  \citenamefont {{Fayer}}, \citenamefont {{Fearon}}, \citenamefont
  {{Fiorucci}}, \citenamefont {{Flaecher}}, \citenamefont {{Ford}},
  \citenamefont {{Francis}}, \citenamefont {{Fraser}}, \citenamefont {{Fruth}},
  \citenamefont {{Gaitskell}}, \citenamefont {{Gantos}}, \citenamefont
  {{Garcia}}, \citenamefont {{Geffre}}, \citenamefont {{Gehman}}, \citenamefont
  {{Genovesi}}, \citenamefont {{Ghag}}, \citenamefont {{Gibbons}},
  \citenamefont {{Gibson}}, \citenamefont {{Gilchriese}}, \citenamefont
  {{Gokhale}}, \citenamefont {{Gomber}}, \citenamefont {{Green}}, \citenamefont
  {{Greenall}}, \citenamefont {{Greenwood}}, \citenamefont {{van der Grinten}},
  \citenamefont {{Gwilliam}}, \citenamefont {{Hall}}, \citenamefont {{Hans}},
  \citenamefont {{Hanzel}}, \citenamefont {{Harrison}}, \citenamefont
  {{Hartigan-O'Connor}}, \citenamefont {{Haselschwardt}}, \citenamefont
  {{Hernandez}}, \citenamefont {{Hertel}}, \citenamefont {{Heuermann}},
  \citenamefont {{Hjemfelt}}, \citenamefont {{Hoff}}, \citenamefont {{Holtom}},
  \citenamefont {{Hor}}, \citenamefont {{Horn}}, \citenamefont {{Huang}},
  \citenamefont {{Hunt}}, \citenamefont {{Ignarra}}, \citenamefont
  {{Jacobsen}}, \citenamefont {{Jahangir}}, \citenamefont {{James}},
  \citenamefont {{Jeffery}}, \citenamefont {{Ji}}, \citenamefont {{Johnson}},
  \citenamefont {{Kaboth}}, \citenamefont {{Kamaha}}, \citenamefont {{Kamdin}},
  \citenamefont {{Kasey}}, \citenamefont {{Kazkaz}}, \citenamefont {{Keefner}},
  \citenamefont {{Khaitan}}, \citenamefont {{Khaleeq}}, \citenamefont
  {{Khazov}}, \citenamefont {{Khurana}}, \citenamefont {{Kim}}, \citenamefont
  {{Kocher}}, \citenamefont {{Kodroff}}, \citenamefont {{Korley}},
  \citenamefont {{Korolkova}}, \citenamefont {{Kras}}, \citenamefont {{Kraus}},
  \citenamefont {{Kravitz}}, \citenamefont {{Krebs}}, \citenamefont
  {{Kreczko}}, \citenamefont {{Krikler}}, \citenamefont {{Kudryavtsev}},
  \citenamefont {{Kyre}}, \citenamefont {{Landerud}}, \citenamefont {{Leason}},
  \citenamefont {{Lee}}, \citenamefont {{Lee}}, \citenamefont {{Leonard}},
  \citenamefont {{Leonard}}, \citenamefont {{Lesko}}, \citenamefont {{Levy}},
  \citenamefont {{Li}}, \citenamefont {{Liao}}, \citenamefont {{Liao}},
  \citenamefont {{Lin}}, \citenamefont {{Lindote}}, \citenamefont {{Linehan}},
  \citenamefont {{Lippincott}}, \citenamefont {{Liu}}, \citenamefont {{Liu}},
  \citenamefont {{Liu}}, \citenamefont {{Loniewski}}, \citenamefont {{Lopes}},
  \citenamefont {{Lopez Asamar}}, \citenamefont {{L{\'o}pez Paredes}},
  \citenamefont {{Lorenzon}}, \citenamefont {{Lucero}}, \citenamefont
  {{Luitz}}, \citenamefont {{Lyle}}, \citenamefont {{Majewski}}, \citenamefont
  {{Makkinje}}, \citenamefont {{Malling}}, \citenamefont {{Manalaysay}},
  \citenamefont {{Manenti}}, \citenamefont {{Mannino}}, \citenamefont
  {{Marangou}}, \citenamefont {{Marzioni}}, \citenamefont {{Maupin}},
  \citenamefont {{McCarthy}}, \citenamefont {{McConnell}}, \citenamefont
  {{McKinsey}}, \citenamefont {{McLaughlin}}, \citenamefont {{Meng}},
  \citenamefont {{Migneault}}, \citenamefont {{Miller}}, \citenamefont
  {{Mizrachi}}, \citenamefont {{Mock}}, \citenamefont {{Monte}}, \citenamefont
  {{Monzani}}, \citenamefont {{Morad}}, \citenamefont {{Morales Mendoza}},
  \citenamefont {{Morrison}}, \citenamefont {{Mount}}, \citenamefont {{Murdy}},
  \citenamefont {{Murphy}}, \citenamefont {{Naim}}, \citenamefont {{Naylor}},
  \citenamefont {{Nedlik}}, \citenamefont {{Nehrkorn}}, \citenamefont
  {{Neves}}, \citenamefont {{Nguyen}}, \citenamefont {{Nikoleyczik}},
  \citenamefont {{Nilima}}, \citenamefont {{O'Dell}}, \citenamefont
  {{O'Neill}}, \citenamefont {{O'Sullivan}}, \citenamefont {{Olcina}},
  \citenamefont {{Olevitch}}, \citenamefont {{Oliver-Mallory}}, \citenamefont
  {{Orpwood}}, \citenamefont {{Pagenkopf}}, \citenamefont {{Pal}},
  \citenamefont {{Palladino}}, \citenamefont {{Palmer}}, \citenamefont
  {{Pangilinan}}, \citenamefont {{Parveen}}, \citenamefont {{Patton}},
  \citenamefont {{Pease}}, \citenamefont {{Penning}}, \citenamefont
  {{Pereira}}, \citenamefont {{Pereira}}, \citenamefont {{Perry}},
  \citenamefont {{Pershing}}, \citenamefont {{Peterson}}, \citenamefont
  {{Piepke}}, \citenamefont {{Podczerwinski}}, \citenamefont {{Porzio}},
  \citenamefont {{Powell}}, \citenamefont {{Preece}}, \citenamefont
  {{Pushkin}}, \citenamefont {{Qie}}, \citenamefont {{Ratcliff}}, \citenamefont
  {{Reichenbacher}}, \citenamefont {{Reichhart}}, \citenamefont {{Rhyne}},
  \citenamefont {{Richards}}, \citenamefont {{Riffard}}, \citenamefont
  {{Rischbieter}}, \citenamefont {{Rodrigues}}, \citenamefont {{Rodriguez}},
  \citenamefont {{Rose}}, \citenamefont {{Rosero}}, \citenamefont {{Rossiter}},
  \citenamefont {{Rushton}}, \citenamefont {{Rutherford}}, \citenamefont
  {{Rynders}}, \citenamefont {{Saba}}, \citenamefont {{Santone}}, \citenamefont
  {{Sazzad}}, \citenamefont {{Schnee}}, \citenamefont {{Scovell}},
  \citenamefont {{Seymour}}, \citenamefont {{Shaw}}, \citenamefont {{Shutt}},
  \citenamefont {{Silk}}, \citenamefont {{Silva}}, \citenamefont {{Sinev}},
  \citenamefont {{Skarpaas}}, \citenamefont {{Skulski}}, \citenamefont
  {{Smith}}, \citenamefont {{Solmaz}}, \citenamefont {{Solovov}}, \citenamefont
  {{Sorensen}}, \citenamefont {{Soria}}, \citenamefont {{Stancu}},
  \citenamefont {{Stark}}, \citenamefont {{Stevens}}, \citenamefont
  {{Stiegler}}, \citenamefont {{Stifter}}, \citenamefont {{Studley}},
  \citenamefont {{Suerfu}}, \citenamefont {{Sumner}}, \citenamefont
  {{Sutcliffe}}, \citenamefont {{Swanson}}, \citenamefont {{Szydagis}},
  \citenamefont {{Tan}}, \citenamefont {{Taylor}}, \citenamefont {{Taylor}},
  \citenamefont {{Taylor}}, \citenamefont {{Temples}}, \citenamefont
  {{Tennyson}}, \citenamefont {{Terman}}, \citenamefont {{Thomas}},
  \citenamefont {{Tiedt}}, \citenamefont {{Timalsina}}, \citenamefont {{To}},
  \citenamefont {{Tom{\'a}s}}, \citenamefont {{Tong}}, \citenamefont {{Tovey}},
  \citenamefont {{Tranter}}, \citenamefont {{Trask}}, \citenamefont
  {{Tripathi}}, \citenamefont {{Tronstad}}, \citenamefont {{Tull}},
  \citenamefont {{Turner}}, \citenamefont {{Tvrznikova}}, \citenamefont
  {{Utku}}, \citenamefont {{Va'Vra}}, \citenamefont {{Vacheret}}, \citenamefont
  {{Vaitkus}}, \citenamefont {{Verbus}}, \citenamefont {{Voirin}},
  \citenamefont {{Waldron}}, \citenamefont {{Wang}}, \citenamefont {{Wang}},
  \citenamefont {{Wang}}, \citenamefont {{Wang}}, \citenamefont {{Wang}},
  \citenamefont {{Watson}}, \citenamefont {{Webb}}, \citenamefont {{White}},
  \citenamefont {{White}}, \citenamefont {{White}}, \citenamefont {{White}},
  \citenamefont {{Whitis}}, \citenamefont {{Williams}}, \citenamefont
  {{Wisniewski}}, \citenamefont {{Witherell}}, \citenamefont {{Wolfs}},
  \citenamefont {{Wolfs}}, \citenamefont {{Woodford}}, \citenamefont
  {{Woodward}}, \citenamefont {{Worm}}, \citenamefont {{Wright}}, \citenamefont
  {{Xia}}, \citenamefont {{Xiang}}, \citenamefont {{Xiao}}, \citenamefont
  {{Xu}}, \citenamefont {{Yeh}}, \citenamefont {{Yin}}, \citenamefont
  {{Young}}, \citenamefont {{Zarzhitsky}}, \citenamefont {{Zuckerman}},
  \citenamefont {{Zweig}},\ and\ \citenamefont {{Lux-Zeplin
  Collaboration}}}]{LZ2023}%
  \BibitemOpen
  \bibfield  {author} {\bibinfo {author} {\bibfnamefont {J.}~\bibnamefont
  {{Aalbers}}}, \bibinfo {author} {\bibfnamefont {D.~S.}\ \bibnamefont
  {{Akerib}}}, \bibinfo {author} {\bibfnamefont {C.~W.}\ \bibnamefont
  {{Akerlof}}}, \bibinfo {author} {\bibfnamefont {A.~K.}\ \bibnamefont {{Al
  Musalhi}}}, \bibinfo {author} {\bibfnamefont {F.}~\bibnamefont {{Alder}}},
  \bibinfo {author} {\bibfnamefont {A.}~\bibnamefont {{Alqahtani}}}, \bibinfo
  {author} {\bibfnamefont {S.~K.}\ \bibnamefont {{Alsum}}}, \bibinfo {author}
  {\bibfnamefont {C.~S.}\ \bibnamefont {{Amarasinghe}}}, \bibinfo {author}
  {\bibfnamefont {A.}~\bibnamefont {{Ames}}}, \bibinfo {author} {\bibfnamefont
  {T.~J.}\ \bibnamefont {{Anderson}}}, \bibinfo {author} {\bibfnamefont
  {N.}~\bibnamefont {{Angelides}}}, \bibinfo {author} {\bibfnamefont {H.~M.}\
  \bibnamefont {{Ara{\'u}jo}}}, \bibinfo {author} {\bibfnamefont {J.~E.}\
  \bibnamefont {{Armstrong}}}, \bibinfo {author} {\bibfnamefont
  {M.}~\bibnamefont {{Arthurs}}}, \bibinfo {author} {\bibfnamefont
  {S.}~\bibnamefont {{Azadi}}}, \bibinfo {author} {\bibfnamefont {A.~J.}\
  \bibnamefont {{Bailey}}}, \bibinfo {author} {\bibfnamefont {A.}~\bibnamefont
  {{Baker}}}, \bibinfo {author} {\bibfnamefont {J.}~\bibnamefont {{Balajthy}}},
  \bibinfo {author} {\bibfnamefont {S.}~\bibnamefont {{Balashov}}}, \bibinfo
  {author} {\bibfnamefont {J.}~\bibnamefont {{Bang}}}, \bibinfo {author}
  {\bibfnamefont {J.~W.}\ \bibnamefont {{Bargemann}}}, \bibinfo {author}
  {\bibfnamefont {M.~J.}\ \bibnamefont {{Barry}}}, \bibinfo {author}
  {\bibfnamefont {J.}~\bibnamefont {{Barthel}}}, \bibinfo {author}
  {\bibfnamefont {D.}~\bibnamefont {{Bauer}}}, \bibinfo {author} {\bibfnamefont
  {A.}~\bibnamefont {{Baxter}}}, \bibinfo {author} {\bibfnamefont
  {K.}~\bibnamefont {{Beattie}}}, \bibinfo {author} {\bibfnamefont
  {J.}~\bibnamefont {{Belle}}}, \bibinfo {author} {\bibfnamefont
  {P.}~\bibnamefont {{Beltrame}}}, \bibinfo {author} {\bibfnamefont
  {J.}~\bibnamefont {{Bensinger}}}, \bibinfo {author} {\bibfnamefont
  {T.}~\bibnamefont {{Benson}}}, \bibinfo {author} {\bibfnamefont {E.~P.}\
  \bibnamefont {{Bernard}}}, \bibinfo {author} {\bibfnamefont {A.}~\bibnamefont
  {{Bhatti}}}, \bibinfo {author} {\bibfnamefont {A.}~\bibnamefont {{Biekert}}},
  \bibinfo {author} {\bibfnamefont {T.~P.}\ \bibnamefont {{Biesiadzinski}}},
  \bibinfo {author} {\bibfnamefont {H.~J.}\ \bibnamefont {{Birch}}}, \bibinfo
  {author} {\bibfnamefont {B.}~\bibnamefont {{Birrittella}}}, \bibinfo {author}
  {\bibfnamefont {G.~M.}\ \bibnamefont {{Blockinger}}}, \bibinfo {author}
  {\bibfnamefont {K.~E.}\ \bibnamefont {{Boast}}}, \bibinfo {author}
  {\bibfnamefont {B.}~\bibnamefont {{Boxer}}}, \bibinfo {author} {\bibfnamefont
  {R.}~\bibnamefont {{Bramante}}}, \bibinfo {author} {\bibfnamefont {C.~A.~J.}\
  \bibnamefont {{Brew}}}, \bibinfo {author} {\bibfnamefont {P.}~\bibnamefont
  {{Br{\'a}s}}}, \bibinfo {author} {\bibfnamefont {J.~H.}\ \bibnamefont
  {{Buckley}}}, \bibinfo {author} {\bibfnamefont {V.~V.}\ \bibnamefont
  {{Bugaev}}}, \bibinfo {author} {\bibfnamefont {S.}~\bibnamefont {{Burdin}}},
  \bibinfo {author} {\bibfnamefont {J.~K.}\ \bibnamefont {{Busenitz}}},
  \bibinfo {author} {\bibfnamefont {M.}~\bibnamefont {{Buuck}}}, \bibinfo
  {author} {\bibfnamefont {R.}~\bibnamefont {{Cabrita}}}, \bibinfo {author}
  {\bibfnamefont {C.}~\bibnamefont {{Carels}}}, \bibinfo {author}
  {\bibfnamefont {D.~L.}\ \bibnamefont {{Carlsmith}}}, \bibinfo {author}
  {\bibfnamefont {B.}~\bibnamefont {{Carlson}}}, \bibinfo {author}
  {\bibfnamefont {M.~C.}\ \bibnamefont {{Carmona-Benitez}}}, \bibinfo {author}
  {\bibfnamefont {M.}~\bibnamefont {{Cascella}}}, \bibinfo {author}
  {\bibfnamefont {C.}~\bibnamefont {{Chan}}}, \bibinfo {author} {\bibfnamefont
  {A.}~\bibnamefont {{Chawla}}}, \bibinfo {author} {\bibfnamefont
  {H.}~\bibnamefont {{Chen}}}, \bibinfo {author} {\bibfnamefont {J.~J.}\
  \bibnamefont {{Cherwinka}}}, \bibinfo {author} {\bibfnamefont {N.~I.}\
  \bibnamefont {{Chott}}}, \bibinfo {author} {\bibfnamefont {A.}~\bibnamefont
  {{Cole}}}, \bibinfo {author} {\bibfnamefont {J.}~\bibnamefont {{Coleman}}},
  \bibinfo {author} {\bibfnamefont {M.~V.}\ \bibnamefont {{Converse}}},
  \bibinfo {author} {\bibfnamefont {A.}~\bibnamefont {{Cottle}}}, \bibinfo
  {author} {\bibfnamefont {G.}~\bibnamefont {{Cox}}}, \bibinfo {author}
  {\bibfnamefont {W.~W.}\ \bibnamefont {{Craddock}}}, \bibinfo {author}
  {\bibfnamefont {O.}~\bibnamefont {{Creaner}}}, \bibinfo {author}
  {\bibfnamefont {D.}~\bibnamefont {{Curran}}}, \bibinfo {author}
  {\bibfnamefont {A.}~\bibnamefont {{Currie}}}, \bibinfo {author}
  {\bibfnamefont {J.~E.}\ \bibnamefont {{Cutter}}}, \bibinfo {author}
  {\bibfnamefont {C.~E.}\ \bibnamefont {{Dahl}}}, \bibinfo {author}
  {\bibfnamefont {A.}~\bibnamefont {{David}}}, \bibinfo {author} {\bibfnamefont
  {J.}~\bibnamefont {{Davis}}}, \bibinfo {author} {\bibfnamefont {T.~J.~R.}\
  \bibnamefont {{Davison}}}, \bibinfo {author} {\bibfnamefont {J.}~\bibnamefont
  {{Delgaudio}}}, \bibinfo {author} {\bibfnamefont {S.}~\bibnamefont {{Dey}}},
  \bibinfo {author} {\bibfnamefont {L.}~\bibnamefont {{de Viveiros}}}, \bibinfo
  {author} {\bibfnamefont {A.}~\bibnamefont {{Dobi}}}, \bibinfo {author}
  {\bibfnamefont {J.~E.~Y.}\ \bibnamefont {{Dobson}}}, \bibinfo {author}
  {\bibfnamefont {E.}~\bibnamefont {{Druszkiewicz}}}, \bibinfo {author}
  {\bibfnamefont {A.}~\bibnamefont {{Dushkin}}}, \bibinfo {author}
  {\bibfnamefont {T.~K.}\ \bibnamefont {{Edberg}}}, \bibinfo {author}
  {\bibfnamefont {W.~R.}\ \bibnamefont {{Edwards}}}, \bibinfo {author}
  {\bibfnamefont {M.~M.}\ \bibnamefont {{Elnimr}}}, \bibinfo {author}
  {\bibfnamefont {W.~T.}\ \bibnamefont {{Emmet}}}, \bibinfo {author}
  {\bibfnamefont {S.~R.}\ \bibnamefont {{Eriksen}}}, \bibinfo {author}
  {\bibfnamefont {C.~H.}\ \bibnamefont {{Faham}}}, \bibinfo {author}
  {\bibfnamefont {A.}~\bibnamefont {{Fan}}}, \bibinfo {author} {\bibfnamefont
  {S.}~\bibnamefont {{Fayer}}}, \bibinfo {author} {\bibfnamefont {N.~M.}\
  \bibnamefont {{Fearon}}}, \bibinfo {author} {\bibfnamefont {S.}~\bibnamefont
  {{Fiorucci}}}, \bibinfo {author} {\bibfnamefont {H.}~\bibnamefont
  {{Flaecher}}}, \bibinfo {author} {\bibfnamefont {P.}~\bibnamefont {{Ford}}},
  \bibinfo {author} {\bibfnamefont {V.~B.}\ \bibnamefont {{Francis}}}, \bibinfo
  {author} {\bibfnamefont {E.~D.}\ \bibnamefont {{Fraser}}}, \bibinfo {author}
  {\bibfnamefont {T.}~\bibnamefont {{Fruth}}}, \bibinfo {author} {\bibfnamefont
  {R.~J.}\ \bibnamefont {{Gaitskell}}}, \bibinfo {author} {\bibfnamefont
  {N.~J.}\ \bibnamefont {{Gantos}}}, \bibinfo {author} {\bibfnamefont
  {D.}~\bibnamefont {{Garcia}}}, \bibinfo {author} {\bibfnamefont
  {A.}~\bibnamefont {{Geffre}}}, \bibinfo {author} {\bibfnamefont {V.~M.}\
  \bibnamefont {{Gehman}}}, \bibinfo {author} {\bibfnamefont {J.}~\bibnamefont
  {{Genovesi}}}, \bibinfo {author} {\bibfnamefont {C.}~\bibnamefont {{Ghag}}},
  \bibinfo {author} {\bibfnamefont {R.}~\bibnamefont {{Gibbons}}}, \bibinfo
  {author} {\bibfnamefont {E.}~\bibnamefont {{Gibson}}}, \bibinfo {author}
  {\bibfnamefont {M.~G.~D.}\ \bibnamefont {{Gilchriese}}}, \bibinfo {author}
  {\bibfnamefont {S.}~\bibnamefont {{Gokhale}}}, \bibinfo {author}
  {\bibfnamefont {B.}~\bibnamefont {{Gomber}}}, \bibinfo {author}
  {\bibfnamefont {J.}~\bibnamefont {{Green}}}, \bibinfo {author} {\bibfnamefont
  {A.}~\bibnamefont {{Greenall}}}, \bibinfo {author} {\bibfnamefont
  {S.}~\bibnamefont {{Greenwood}}}, \bibinfo {author} {\bibfnamefont
  {M.~G.~D.}\ \bibnamefont {{van der Grinten}}}, \bibinfo {author}
  {\bibfnamefont {C.~B.}\ \bibnamefont {{Gwilliam}}}, \bibinfo {author}
  {\bibfnamefont {C.~R.}\ \bibnamefont {{Hall}}}, \bibinfo {author}
  {\bibfnamefont {S.}~\bibnamefont {{Hans}}}, \bibinfo {author} {\bibfnamefont
  {K.}~\bibnamefont {{Hanzel}}}, \bibinfo {author} {\bibfnamefont
  {A.}~\bibnamefont {{Harrison}}}, \bibinfo {author} {\bibfnamefont
  {E.}~\bibnamefont {{Hartigan-O'Connor}}}, \bibinfo {author} {\bibfnamefont
  {S.~J.}\ \bibnamefont {{Haselschwardt}}}, \bibinfo {author} {\bibfnamefont
  {M.~A.}\ \bibnamefont {{Hernandez}}}, \bibinfo {author} {\bibfnamefont
  {S.~A.}\ \bibnamefont {{Hertel}}}, \bibinfo {author} {\bibfnamefont
  {G.}~\bibnamefont {{Heuermann}}}, \bibinfo {author} {\bibfnamefont
  {C.}~\bibnamefont {{Hjemfelt}}}, \bibinfo {author} {\bibfnamefont {M.~D.}\
  \bibnamefont {{Hoff}}}, \bibinfo {author} {\bibfnamefont {E.}~\bibnamefont
  {{Holtom}}}, \bibinfo {author} {\bibfnamefont {J.~Y.~K.}\ \bibnamefont
  {{Hor}}}, \bibinfo {author} {\bibfnamefont {M.}~\bibnamefont {{Horn}}},
  \bibinfo {author} {\bibfnamefont {D.~Q.}\ \bibnamefont {{Huang}}}, \bibinfo
  {author} {\bibfnamefont {D.}~\bibnamefont {{Hunt}}}, \bibinfo {author}
  {\bibfnamefont {C.~M.}\ \bibnamefont {{Ignarra}}}, \bibinfo {author}
  {\bibfnamefont {R.~G.}\ \bibnamefont {{Jacobsen}}}, \bibinfo {author}
  {\bibfnamefont {O.}~\bibnamefont {{Jahangir}}}, \bibinfo {author}
  {\bibfnamefont {R.~S.}\ \bibnamefont {{James}}}, \bibinfo {author}
  {\bibfnamefont {S.~N.}\ \bibnamefont {{Jeffery}}}, \bibinfo {author}
  {\bibfnamefont {W.}~\bibnamefont {{Ji}}}, \bibinfo {author} {\bibfnamefont
  {J.}~\bibnamefont {{Johnson}}}, \bibinfo {author} {\bibfnamefont {A.~C.}\
  \bibnamefont {{Kaboth}}}, \bibinfo {author} {\bibfnamefont {A.~C.}\
  \bibnamefont {{Kamaha}}}, \bibinfo {author} {\bibfnamefont {K.}~\bibnamefont
  {{Kamdin}}}, \bibinfo {author} {\bibfnamefont {V.}~\bibnamefont {{Kasey}}},
  \bibinfo {author} {\bibfnamefont {K.}~\bibnamefont {{Kazkaz}}}, \bibinfo
  {author} {\bibfnamefont {J.}~\bibnamefont {{Keefner}}}, \bibinfo {author}
  {\bibfnamefont {D.}~\bibnamefont {{Khaitan}}}, \bibinfo {author}
  {\bibfnamefont {M.}~\bibnamefont {{Khaleeq}}}, \bibinfo {author}
  {\bibfnamefont {A.}~\bibnamefont {{Khazov}}}, \bibinfo {author}
  {\bibfnamefont {I.}~\bibnamefont {{Khurana}}}, \bibinfo {author}
  {\bibfnamefont {Y.~D.}\ \bibnamefont {{Kim}}}, \bibinfo {author}
  {\bibfnamefont {C.~D.}\ \bibnamefont {{Kocher}}}, \bibinfo {author}
  {\bibfnamefont {D.}~\bibnamefont {{Kodroff}}}, \bibinfo {author}
  {\bibfnamefont {L.}~\bibnamefont {{Korley}}}, \bibinfo {author}
  {\bibfnamefont {E.~V.}\ \bibnamefont {{Korolkova}}}, \bibinfo {author}
  {\bibfnamefont {J.}~\bibnamefont {{Kras}}}, \bibinfo {author} {\bibfnamefont
  {H.}~\bibnamefont {{Kraus}}}, \bibinfo {author} {\bibfnamefont
  {S.}~\bibnamefont {{Kravitz}}}, \bibinfo {author} {\bibfnamefont {H.~J.}\
  \bibnamefont {{Krebs}}}, \bibinfo {author} {\bibfnamefont {L.}~\bibnamefont
  {{Kreczko}}}, \bibinfo {author} {\bibfnamefont {B.}~\bibnamefont
  {{Krikler}}}, \bibinfo {author} {\bibfnamefont {V.~A.}\ \bibnamefont
  {{Kudryavtsev}}}, \bibinfo {author} {\bibfnamefont {S.}~\bibnamefont
  {{Kyre}}}, \bibinfo {author} {\bibfnamefont {B.}~\bibnamefont {{Landerud}}},
  \bibinfo {author} {\bibfnamefont {E.~A.}\ \bibnamefont {{Leason}}}, \bibinfo
  {author} {\bibfnamefont {C.}~\bibnamefont {{Lee}}}, \bibinfo {author}
  {\bibfnamefont {J.}~\bibnamefont {{Lee}}}, \bibinfo {author} {\bibfnamefont
  {D.~S.}\ \bibnamefont {{Leonard}}}, \bibinfo {author} {\bibfnamefont
  {R.}~\bibnamefont {{Leonard}}}, \bibinfo {author} {\bibfnamefont {K.~T.}\
  \bibnamefont {{Lesko}}}, \bibinfo {author} {\bibfnamefont {C.}~\bibnamefont
  {{Levy}}}, \bibinfo {author} {\bibfnamefont {J.}~\bibnamefont {{Li}}},
  \bibinfo {author} {\bibfnamefont {F.~T.}\ \bibnamefont {{Liao}}}, \bibinfo
  {author} {\bibfnamefont {J.}~\bibnamefont {{Liao}}}, \bibinfo {author}
  {\bibfnamefont {J.}~\bibnamefont {{Lin}}}, \bibinfo {author} {\bibfnamefont
  {A.}~\bibnamefont {{Lindote}}}, \bibinfo {author} {\bibfnamefont
  {R.}~\bibnamefont {{Linehan}}}, \bibinfo {author} {\bibfnamefont {W.~H.}\
  \bibnamefont {{Lippincott}}}, \bibinfo {author} {\bibfnamefont
  {R.}~\bibnamefont {{Liu}}}, \bibinfo {author} {\bibfnamefont
  {X.}~\bibnamefont {{Liu}}}, \bibinfo {author} {\bibfnamefont
  {Y.}~\bibnamefont {{Liu}}}, \bibinfo {author} {\bibfnamefont
  {C.}~\bibnamefont {{Loniewski}}}, \bibinfo {author} {\bibfnamefont {M.~I.}\
  \bibnamefont {{Lopes}}}, \bibinfo {author} {\bibfnamefont {E.}~\bibnamefont
  {{Lopez Asamar}}}, \bibinfo {author} {\bibfnamefont {B.}~\bibnamefont
  {{L{\'o}pez Paredes}}}, \bibinfo {author} {\bibfnamefont {W.}~\bibnamefont
  {{Lorenzon}}}, \bibinfo {author} {\bibfnamefont {D.}~\bibnamefont
  {{Lucero}}}, \bibinfo {author} {\bibfnamefont {S.}~\bibnamefont {{Luitz}}},
  \bibinfo {author} {\bibfnamefont {J.~M.}\ \bibnamefont {{Lyle}}}, \bibinfo
  {author} {\bibfnamefont {P.~A.}\ \bibnamefont {{Majewski}}}, \bibinfo
  {author} {\bibfnamefont {J.}~\bibnamefont {{Makkinje}}}, \bibinfo {author}
  {\bibfnamefont {D.~C.}\ \bibnamefont {{Malling}}}, \bibinfo {author}
  {\bibfnamefont {A.}~\bibnamefont {{Manalaysay}}}, \bibinfo {author}
  {\bibfnamefont {L.}~\bibnamefont {{Manenti}}}, \bibinfo {author}
  {\bibfnamefont {R.~L.}\ \bibnamefont {{Mannino}}}, \bibinfo {author}
  {\bibfnamefont {N.}~\bibnamefont {{Marangou}}}, \bibinfo {author}
  {\bibfnamefont {M.~F.}\ \bibnamefont {{Marzioni}}}, \bibinfo {author}
  {\bibfnamefont {C.}~\bibnamefont {{Maupin}}}, \bibinfo {author}
  {\bibfnamefont {M.~E.}\ \bibnamefont {{McCarthy}}}, \bibinfo {author}
  {\bibfnamefont {C.~T.}\ \bibnamefont {{McConnell}}}, \bibinfo {author}
  {\bibfnamefont {D.~N.}\ \bibnamefont {{McKinsey}}}, \bibinfo {author}
  {\bibfnamefont {J.}~\bibnamefont {{McLaughlin}}}, \bibinfo {author}
  {\bibfnamefont {Y.}~\bibnamefont {{Meng}}}, \bibinfo {author} {\bibfnamefont
  {J.}~\bibnamefont {{Migneault}}}, \bibinfo {author} {\bibfnamefont {E.~H.}\
  \bibnamefont {{Miller}}}, \bibinfo {author} {\bibfnamefont {E.}~\bibnamefont
  {{Mizrachi}}}, \bibinfo {author} {\bibfnamefont {J.~A.}\ \bibnamefont
  {{Mock}}}, \bibinfo {author} {\bibfnamefont {A.}~\bibnamefont {{Monte}}},
  \bibinfo {author} {\bibfnamefont {M.~E.}\ \bibnamefont {{Monzani}}}, \bibinfo
  {author} {\bibfnamefont {J.~A.}\ \bibnamefont {{Morad}}}, \bibinfo {author}
  {\bibfnamefont {J.~D.}\ \bibnamefont {{Morales Mendoza}}}, \bibinfo {author}
  {\bibfnamefont {E.}~\bibnamefont {{Morrison}}}, \bibinfo {author}
  {\bibfnamefont {B.~J.}\ \bibnamefont {{Mount}}}, \bibinfo {author}
  {\bibfnamefont {M.}~\bibnamefont {{Murdy}}}, \bibinfo {author} {\bibfnamefont
  {A.~S.~J.}\ \bibnamefont {{Murphy}}}, \bibinfo {author} {\bibfnamefont
  {D.}~\bibnamefont {{Naim}}}, \bibinfo {author} {\bibfnamefont
  {A.}~\bibnamefont {{Naylor}}}, \bibinfo {author} {\bibfnamefont
  {C.}~\bibnamefont {{Nedlik}}}, \bibinfo {author} {\bibfnamefont
  {C.}~\bibnamefont {{Nehrkorn}}}, \bibinfo {author} {\bibfnamefont
  {F.}~\bibnamefont {{Neves}}}, \bibinfo {author} {\bibfnamefont
  {A.}~\bibnamefont {{Nguyen}}}, \bibinfo {author} {\bibfnamefont {J.~A.}\
  \bibnamefont {{Nikoleyczik}}}, \bibinfo {author} {\bibfnamefont
  {A.}~\bibnamefont {{Nilima}}}, \bibinfo {author} {\bibfnamefont
  {J.}~\bibnamefont {{O'Dell}}}, \bibinfo {author} {\bibfnamefont {F.~G.}\
  \bibnamefont {{O'Neill}}}, \bibinfo {author} {\bibfnamefont {K.}~\bibnamefont
  {{O'Sullivan}}}, \bibinfo {author} {\bibfnamefont {I.}~\bibnamefont
  {{Olcina}}}, \bibinfo {author} {\bibfnamefont {M.~A.}\ \bibnamefont
  {{Olevitch}}}, \bibinfo {author} {\bibfnamefont {K.~C.}\ \bibnamefont
  {{Oliver-Mallory}}}, \bibinfo {author} {\bibfnamefont {J.}~\bibnamefont
  {{Orpwood}}}, \bibinfo {author} {\bibfnamefont {D.}~\bibnamefont
  {{Pagenkopf}}}, \bibinfo {author} {\bibfnamefont {S.}~\bibnamefont {{Pal}}},
  \bibinfo {author} {\bibfnamefont {K.~J.}\ \bibnamefont {{Palladino}}},
  \bibinfo {author} {\bibfnamefont {J.}~\bibnamefont {{Palmer}}}, \bibinfo
  {author} {\bibfnamefont {M.}~\bibnamefont {{Pangilinan}}}, \bibinfo {author}
  {\bibfnamefont {N.}~\bibnamefont {{Parveen}}}, \bibinfo {author}
  {\bibfnamefont {S.~J.}\ \bibnamefont {{Patton}}}, \bibinfo {author}
  {\bibfnamefont {E.~K.}\ \bibnamefont {{Pease}}}, \bibinfo {author}
  {\bibfnamefont {B.}~\bibnamefont {{Penning}}}, \bibinfo {author}
  {\bibfnamefont {C.}~\bibnamefont {{Pereira}}}, \bibinfo {author}
  {\bibfnamefont {G.}~\bibnamefont {{Pereira}}}, \bibinfo {author}
  {\bibfnamefont {E.}~\bibnamefont {{Perry}}}, \bibinfo {author} {\bibfnamefont
  {T.}~\bibnamefont {{Pershing}}}, \bibinfo {author} {\bibfnamefont {I.~B.}\
  \bibnamefont {{Peterson}}}, \bibinfo {author} {\bibfnamefont
  {A.}~\bibnamefont {{Piepke}}}, \bibinfo {author} {\bibfnamefont
  {J.}~\bibnamefont {{Podczerwinski}}}, \bibinfo {author} {\bibfnamefont
  {D.}~\bibnamefont {{Porzio}}}, \bibinfo {author} {\bibfnamefont
  {S.}~\bibnamefont {{Powell}}}, \bibinfo {author} {\bibfnamefont {R.~M.}\
  \bibnamefont {{Preece}}}, \bibinfo {author} {\bibfnamefont {K.}~\bibnamefont
  {{Pushkin}}}, \bibinfo {author} {\bibfnamefont {Y.}~\bibnamefont {{Qie}}},
  \bibinfo {author} {\bibfnamefont {B.~N.}\ \bibnamefont {{Ratcliff}}},
  \bibinfo {author} {\bibfnamefont {J.}~\bibnamefont {{Reichenbacher}}},
  \bibinfo {author} {\bibfnamefont {L.}~\bibnamefont {{Reichhart}}}, \bibinfo
  {author} {\bibfnamefont {C.~A.}\ \bibnamefont {{Rhyne}}}, \bibinfo {author}
  {\bibfnamefont {A.}~\bibnamefont {{Richards}}}, \bibinfo {author}
  {\bibfnamefont {Q.}~\bibnamefont {{Riffard}}}, \bibinfo {author}
  {\bibfnamefont {G.~R.~C.}\ \bibnamefont {{Rischbieter}}}, \bibinfo {author}
  {\bibfnamefont {J.~P.}\ \bibnamefont {{Rodrigues}}}, \bibinfo {author}
  {\bibfnamefont {A.}~\bibnamefont {{Rodriguez}}}, \bibinfo {author}
  {\bibfnamefont {H.~J.}\ \bibnamefont {{Rose}}}, \bibinfo {author}
  {\bibfnamefont {R.}~\bibnamefont {{Rosero}}}, \bibinfo {author}
  {\bibfnamefont {P.}~\bibnamefont {{Rossiter}}}, \bibinfo {author}
  {\bibfnamefont {T.}~\bibnamefont {{Rushton}}}, \bibinfo {author}
  {\bibfnamefont {G.}~\bibnamefont {{Rutherford}}}, \bibinfo {author}
  {\bibfnamefont {D.}~\bibnamefont {{Rynders}}}, \bibinfo {author}
  {\bibfnamefont {J.~S.}\ \bibnamefont {{Saba}}}, \bibinfo {author}
  {\bibfnamefont {D.}~\bibnamefont {{Santone}}}, \bibinfo {author}
  {\bibfnamefont {A.~B.~M.~R.}\ \bibnamefont {{Sazzad}}}, \bibinfo {author}
  {\bibfnamefont {R.~W.}\ \bibnamefont {{Schnee}}}, \bibinfo {author}
  {\bibfnamefont {P.~R.}\ \bibnamefont {{Scovell}}}, \bibinfo {author}
  {\bibfnamefont {D.}~\bibnamefont {{Seymour}}}, \bibinfo {author}
  {\bibfnamefont {S.}~\bibnamefont {{Shaw}}}, \bibinfo {author} {\bibfnamefont
  {T.}~\bibnamefont {{Shutt}}}, \bibinfo {author} {\bibfnamefont {J.~J.}\
  \bibnamefont {{Silk}}}, \bibinfo {author} {\bibfnamefont {C.}~\bibnamefont
  {{Silva}}}, \bibinfo {author} {\bibfnamefont {G.}~\bibnamefont {{Sinev}}},
  \bibinfo {author} {\bibfnamefont {K.}~\bibnamefont {{Skarpaas}}}, \bibinfo
  {author} {\bibfnamefont {W.}~\bibnamefont {{Skulski}}}, \bibinfo {author}
  {\bibfnamefont {R.}~\bibnamefont {{Smith}}}, \bibinfo {author} {\bibfnamefont
  {M.}~\bibnamefont {{Solmaz}}}, \bibinfo {author} {\bibfnamefont {V.~N.}\
  \bibnamefont {{Solovov}}}, \bibinfo {author} {\bibfnamefont {P.}~\bibnamefont
  {{Sorensen}}}, \bibinfo {author} {\bibfnamefont {J.}~\bibnamefont {{Soria}}},
  \bibinfo {author} {\bibfnamefont {I.}~\bibnamefont {{Stancu}}}, \bibinfo
  {author} {\bibfnamefont {M.~R.}\ \bibnamefont {{Stark}}}, \bibinfo {author}
  {\bibfnamefont {A.}~\bibnamefont {{Stevens}}}, \bibinfo {author}
  {\bibfnamefont {T.~M.}\ \bibnamefont {{Stiegler}}}, \bibinfo {author}
  {\bibfnamefont {K.}~\bibnamefont {{Stifter}}}, \bibinfo {author}
  {\bibfnamefont {R.}~\bibnamefont {{Studley}}}, \bibinfo {author}
  {\bibfnamefont {B.}~\bibnamefont {{Suerfu}}}, \bibinfo {author}
  {\bibfnamefont {T.~J.}\ \bibnamefont {{Sumner}}}, \bibinfo {author}
  {\bibfnamefont {P.}~\bibnamefont {{Sutcliffe}}}, \bibinfo {author}
  {\bibfnamefont {N.}~\bibnamefont {{Swanson}}}, \bibinfo {author}
  {\bibfnamefont {M.}~\bibnamefont {{Szydagis}}}, \bibinfo {author}
  {\bibfnamefont {M.}~\bibnamefont {{Tan}}}, \bibinfo {author} {\bibfnamefont
  {D.~J.}\ \bibnamefont {{Taylor}}}, \bibinfo {author} {\bibfnamefont
  {R.}~\bibnamefont {{Taylor}}}, \bibinfo {author} {\bibfnamefont {W.~C.}\
  \bibnamefont {{Taylor}}}, \bibinfo {author} {\bibfnamefont {D.~J.}\
  \bibnamefont {{Temples}}}, \bibinfo {author} {\bibfnamefont {B.~P.}\
  \bibnamefont {{Tennyson}}}, \bibinfo {author} {\bibfnamefont {P.~A.}\
  \bibnamefont {{Terman}}}, \bibinfo {author} {\bibfnamefont {K.~J.}\
  \bibnamefont {{Thomas}}}, \bibinfo {author} {\bibfnamefont {D.~R.}\
  \bibnamefont {{Tiedt}}}, \bibinfo {author} {\bibfnamefont {M.}~\bibnamefont
  {{Timalsina}}}, \bibinfo {author} {\bibfnamefont {W.~H.}\ \bibnamefont
  {{To}}}, \bibinfo {author} {\bibfnamefont {A.}~\bibnamefont {{Tom{\'a}s}}},
  \bibinfo {author} {\bibfnamefont {Z.}~\bibnamefont {{Tong}}}, \bibinfo
  {author} {\bibfnamefont {D.~R.}\ \bibnamefont {{Tovey}}}, \bibinfo {author}
  {\bibfnamefont {J.}~\bibnamefont {{Tranter}}}, \bibinfo {author}
  {\bibfnamefont {M.}~\bibnamefont {{Trask}}}, \bibinfo {author} {\bibfnamefont
  {M.}~\bibnamefont {{Tripathi}}}, \bibinfo {author} {\bibfnamefont {D.~R.}\
  \bibnamefont {{Tronstad}}}, \bibinfo {author} {\bibfnamefont {C.~E.}\
  \bibnamefont {{Tull}}}, \bibinfo {author} {\bibfnamefont {W.}~\bibnamefont
  {{Turner}}}, \bibinfo {author} {\bibfnamefont {L.}~\bibnamefont
  {{Tvrznikova}}}, \bibinfo {author} {\bibfnamefont {U.}~\bibnamefont
  {{Utku}}}, \bibinfo {author} {\bibfnamefont {J.}~\bibnamefont {{Va'Vra}}},
  \bibinfo {author} {\bibfnamefont {A.}~\bibnamefont {{Vacheret}}}, \bibinfo
  {author} {\bibfnamefont {A.~C.}\ \bibnamefont {{Vaitkus}}}, \bibinfo {author}
  {\bibfnamefont {J.~R.}\ \bibnamefont {{Verbus}}}, \bibinfo {author}
  {\bibfnamefont {E.}~\bibnamefont {{Voirin}}}, \bibinfo {author}
  {\bibfnamefont {W.~L.}\ \bibnamefont {{Waldron}}}, \bibinfo {author}
  {\bibfnamefont {A.}~\bibnamefont {{Wang}}}, \bibinfo {author} {\bibfnamefont
  {B.}~\bibnamefont {{Wang}}}, \bibinfo {author} {\bibfnamefont {J.~J.}\
  \bibnamefont {{Wang}}}, \bibinfo {author} {\bibfnamefont {W.}~\bibnamefont
  {{Wang}}}, \bibinfo {author} {\bibfnamefont {Y.}~\bibnamefont {{Wang}}},
  \bibinfo {author} {\bibfnamefont {J.~R.}\ \bibnamefont {{Watson}}}, \bibinfo
  {author} {\bibfnamefont {R.~C.}\ \bibnamefont {{Webb}}}, \bibinfo {author}
  {\bibfnamefont {A.}~\bibnamefont {{White}}}, \bibinfo {author} {\bibfnamefont
  {D.~T.}\ \bibnamefont {{White}}}, \bibinfo {author} {\bibfnamefont {J.~T.}\
  \bibnamefont {{White}}}, \bibinfo {author} {\bibfnamefont {R.~G.}\
  \bibnamefont {{White}}}, \bibinfo {author} {\bibfnamefont {T.~J.}\
  \bibnamefont {{Whitis}}}, \bibinfo {author} {\bibfnamefont {M.}~\bibnamefont
  {{Williams}}}, \bibinfo {author} {\bibfnamefont {W.~J.}\ \bibnamefont
  {{Wisniewski}}}, \bibinfo {author} {\bibfnamefont {M.~S.}\ \bibnamefont
  {{Witherell}}}, \bibinfo {author} {\bibfnamefont {F.~L.~H.}\ \bibnamefont
  {{Wolfs}}}, \bibinfo {author} {\bibfnamefont {J.~D.}\ \bibnamefont
  {{Wolfs}}}, \bibinfo {author} {\bibfnamefont {S.}~\bibnamefont {{Woodford}}},
  \bibinfo {author} {\bibfnamefont {D.}~\bibnamefont {{Woodward}}}, \bibinfo
  {author} {\bibfnamefont {S.~D.}\ \bibnamefont {{Worm}}}, \bibinfo {author}
  {\bibfnamefont {C.~J.}\ \bibnamefont {{Wright}}}, \bibinfo {author}
  {\bibfnamefont {Q.}~\bibnamefont {{Xia}}}, \bibinfo {author} {\bibfnamefont
  {X.}~\bibnamefont {{Xiang}}}, \bibinfo {author} {\bibfnamefont
  {Q.}~\bibnamefont {{Xiao}}}, \bibinfo {author} {\bibfnamefont
  {J.}~\bibnamefont {{Xu}}}, \bibinfo {author} {\bibfnamefont {M.}~\bibnamefont
  {{Yeh}}}, \bibinfo {author} {\bibfnamefont {J.}~\bibnamefont {{Yin}}},
  \bibinfo {author} {\bibfnamefont {I.}~\bibnamefont {{Young}}}, \bibinfo
  {author} {\bibfnamefont {P.}~\bibnamefont {{Zarzhitsky}}}, \bibinfo {author}
  {\bibfnamefont {A.}~\bibnamefont {{Zuckerman}}}, \bibinfo {author}
  {\bibfnamefont {E.~A.}\ \bibnamefont {{Zweig}}},\ and\ \bibinfo {author}
  {\bibnamefont {{Lux-Zeplin Collaboration}}},\ }\bibfield  {title} {\bibinfo
  {title} {{First Dark Matter Search Results from the LUX-ZEPLIN (LZ)
  Experiment}},\ }\href {https://doi.org/10.1103/PhysRevLett.131.041002}
  {\bibfield  {journal} {\bibinfo  {journal} {\prl}\ }\textbf {\bibinfo
  {volume} {131}},\ \bibinfo {eid} {041002} (\bibinfo {year} {2023})},\ \Eprint
  {https://arxiv.org/abs/2207.03764} {arXiv:2207.03764 [hep-ex]} \BibitemShut
  {NoStop}%
\bibitem [{\citenamefont {{Li}}\ \emph {et~al.}(2023)\citenamefont {{Li}},
  \citenamefont {{Wu}}, \citenamefont {{Abdukerim}}, \citenamefont {{Bo}},
  \citenamefont {{Chen}}, \citenamefont {{Chen}}, \citenamefont {{Chen}},
  \citenamefont {{Cheng}}, \citenamefont {{Cheng}}, \citenamefont {{Cui}},
  \citenamefont {{Fan}}, \citenamefont {{Fang}}, \citenamefont {{Fu}},
  \citenamefont {{Fu}}, \citenamefont {{Geng}}, \citenamefont {{Giboni}},
  \citenamefont {{Gu}}, \citenamefont {{Guo}}, \citenamefont {{Han}},
  \citenamefont {{Han}}, \citenamefont {{He}}, \citenamefont {{He}},
  \citenamefont {{Huang}}, \citenamefont {{Huang}}, \citenamefont {{Huang}},
  \citenamefont {{Hou}}, \citenamefont {{Ji}}, \citenamefont {{Ju}},
  \citenamefont {{Li}}, \citenamefont {{Li}}, \citenamefont {{Li}},
  \citenamefont {{Li}}, \citenamefont {{Lin}}, \citenamefont {{Liu}},
  \citenamefont {{Lu}}, \citenamefont {{Luo}}, \citenamefont {{Luo}},
  \citenamefont {{Ma}}, \citenamefont {{Ma}}, \citenamefont {{Mao}},
  \citenamefont {{Meng}}, \citenamefont {{Ning}}, \citenamefont {{Qi}},
  \citenamefont {{Qian}}, \citenamefont {{Ren}}, \citenamefont {{Shaheed}},
  \citenamefont {{Shang}}, \citenamefont {{Shang}}, \citenamefont {{Shen}},
  \citenamefont {{Si}}, \citenamefont {{Sun}}, \citenamefont {{Tan}},
  \citenamefont {{Tao}}, \citenamefont {{Wang}}, \citenamefont {{Wang}},
  \citenamefont {{Wang}}, \citenamefont {{Wang}}, \citenamefont {{Wang}},
  \citenamefont {{Wang}}, \citenamefont {{Wang}}, \citenamefont {{Wang}},
  \citenamefont {{Wei}}, \citenamefont {{Wu}}, \citenamefont {{Xia}},
  \citenamefont {{Xiao}}, \citenamefont {{Xiao}}, \citenamefont {{Xie}},
  \citenamefont {{Yan}}, \citenamefont {{Yan}}, \citenamefont {{Yang}},
  \citenamefont {{Yang}}, \citenamefont {{Yao}}, \citenamefont {{You}},
  \citenamefont {{Yu}}, \citenamefont {{Yuan}}, \citenamefont {{Yuan}},
  \citenamefont {{Yuan}}, \citenamefont {{Zeng}}, \citenamefont {{Zhang}},
  \citenamefont {{Zhang}}, \citenamefont {{Zhang}}, \citenamefont {{Zhang}},
  \citenamefont {{Zhang}}, \citenamefont {{Zhang}}, \citenamefont {{Zhang}},
  \citenamefont {{Zhang}}, \citenamefont {{Zhang}}, \citenamefont {{Zhao}},
  \citenamefont {{Zheng}}, \citenamefont {{Zhou}}, \citenamefont {{Zhou}},
  \citenamefont {{Zhou}}, \citenamefont {{Zhou}}, \citenamefont {{Zhou}},\ and\
  \citenamefont {{PandaX Collaboration}}}]{PandaX-4T2023}%
  \BibitemOpen
  \bibfield  {author} {\bibinfo {author} {\bibfnamefont {S.}~\bibnamefont
  {{Li}}}, \bibinfo {author} {\bibfnamefont {M.}~\bibnamefont {{Wu}}}, \bibinfo
  {author} {\bibfnamefont {A.}~\bibnamefont {{Abdukerim}}}, \bibinfo {author}
  {\bibfnamefont {Z.}~\bibnamefont {{Bo}}}, \bibinfo {author} {\bibfnamefont
  {W.}~\bibnamefont {{Chen}}}, \bibinfo {author} {\bibfnamefont
  {X.}~\bibnamefont {{Chen}}}, \bibinfo {author} {\bibfnamefont
  {Y.}~\bibnamefont {{Chen}}}, \bibinfo {author} {\bibfnamefont
  {C.}~\bibnamefont {{Cheng}}}, \bibinfo {author} {\bibfnamefont
  {Z.}~\bibnamefont {{Cheng}}}, \bibinfo {author} {\bibfnamefont
  {X.}~\bibnamefont {{Cui}}}, \bibinfo {author} {\bibfnamefont
  {Y.}~\bibnamefont {{Fan}}}, \bibinfo {author} {\bibfnamefont
  {D.}~\bibnamefont {{Fang}}}, \bibinfo {author} {\bibfnamefont
  {C.}~\bibnamefont {{Fu}}}, \bibinfo {author} {\bibfnamefont {M.}~\bibnamefont
  {{Fu}}}, \bibinfo {author} {\bibfnamefont {L.}~\bibnamefont {{Geng}}},
  \bibinfo {author} {\bibfnamefont {K.}~\bibnamefont {{Giboni}}}, \bibinfo
  {author} {\bibfnamefont {L.}~\bibnamefont {{Gu}}}, \bibinfo {author}
  {\bibfnamefont {X.}~\bibnamefont {{Guo}}}, \bibinfo {author} {\bibfnamefont
  {C.}~\bibnamefont {{Han}}}, \bibinfo {author} {\bibfnamefont
  {K.}~\bibnamefont {{Han}}}, \bibinfo {author} {\bibfnamefont
  {C.}~\bibnamefont {{He}}}, \bibinfo {author} {\bibfnamefont {J.}~\bibnamefont
  {{He}}}, \bibinfo {author} {\bibfnamefont {D.}~\bibnamefont {{Huang}}},
  \bibinfo {author} {\bibfnamefont {Y.}~\bibnamefont {{Huang}}}, \bibinfo
  {author} {\bibfnamefont {Z.}~\bibnamefont {{Huang}}}, \bibinfo {author}
  {\bibfnamefont {R.}~\bibnamefont {{Hou}}}, \bibinfo {author} {\bibfnamefont
  {X.}~\bibnamefont {{Ji}}}, \bibinfo {author} {\bibfnamefont {Y.}~\bibnamefont
  {{Ju}}}, \bibinfo {author} {\bibfnamefont {C.}~\bibnamefont {{Li}}}, \bibinfo
  {author} {\bibfnamefont {J.}~\bibnamefont {{Li}}}, \bibinfo {author}
  {\bibfnamefont {M.}~\bibnamefont {{Li}}}, \bibinfo {author} {\bibfnamefont
  {S.}~\bibnamefont {{Li}}}, \bibinfo {author} {\bibfnamefont {Q.}~\bibnamefont
  {{Lin}}}, \bibinfo {author} {\bibfnamefont {J.}~\bibnamefont {{Liu}}},
  \bibinfo {author} {\bibfnamefont {X.}~\bibnamefont {{Lu}}}, \bibinfo {author}
  {\bibfnamefont {L.}~\bibnamefont {{Luo}}}, \bibinfo {author} {\bibfnamefont
  {Y.}~\bibnamefont {{Luo}}}, \bibinfo {author} {\bibfnamefont
  {W.}~\bibnamefont {{Ma}}}, \bibinfo {author} {\bibfnamefont {Y.}~\bibnamefont
  {{Ma}}}, \bibinfo {author} {\bibfnamefont {Y.}~\bibnamefont {{Mao}}},
  \bibinfo {author} {\bibfnamefont {Y.}~\bibnamefont {{Meng}}}, \bibinfo
  {author} {\bibfnamefont {X.}~\bibnamefont {{Ning}}}, \bibinfo {author}
  {\bibfnamefont {N.}~\bibnamefont {{Qi}}}, \bibinfo {author} {\bibfnamefont
  {Z.}~\bibnamefont {{Qian}}}, \bibinfo {author} {\bibfnamefont
  {X.}~\bibnamefont {{Ren}}}, \bibinfo {author} {\bibfnamefont
  {N.}~\bibnamefont {{Shaheed}}}, \bibinfo {author} {\bibfnamefont
  {C.}~\bibnamefont {{Shang}}}, \bibinfo {author} {\bibfnamefont
  {X.}~\bibnamefont {{Shang}}}, \bibinfo {author} {\bibfnamefont
  {G.}~\bibnamefont {{Shen}}}, \bibinfo {author} {\bibfnamefont
  {L.}~\bibnamefont {{Si}}}, \bibinfo {author} {\bibfnamefont {W.}~\bibnamefont
  {{Sun}}}, \bibinfo {author} {\bibfnamefont {A.}~\bibnamefont {{Tan}}},
  \bibinfo {author} {\bibfnamefont {Y.}~\bibnamefont {{Tao}}}, \bibinfo
  {author} {\bibfnamefont {A.}~\bibnamefont {{Wang}}}, \bibinfo {author}
  {\bibfnamefont {M.}~\bibnamefont {{Wang}}}, \bibinfo {author} {\bibfnamefont
  {Q.}~\bibnamefont {{Wang}}}, \bibinfo {author} {\bibfnamefont
  {S.}~\bibnamefont {{Wang}}}, \bibinfo {author} {\bibfnamefont
  {S.}~\bibnamefont {{Wang}}}, \bibinfo {author} {\bibfnamefont
  {W.}~\bibnamefont {{Wang}}}, \bibinfo {author} {\bibfnamefont
  {X.}~\bibnamefont {{Wang}}}, \bibinfo {author} {\bibfnamefont
  {Z.}~\bibnamefont {{Wang}}}, \bibinfo {author} {\bibfnamefont
  {Y.}~\bibnamefont {{Wei}}}, \bibinfo {author} {\bibfnamefont
  {W.}~\bibnamefont {{Wu}}}, \bibinfo {author} {\bibfnamefont {J.}~\bibnamefont
  {{Xia}}}, \bibinfo {author} {\bibfnamefont {M.}~\bibnamefont {{Xiao}}},
  \bibinfo {author} {\bibfnamefont {X.}~\bibnamefont {{Xiao}}}, \bibinfo
  {author} {\bibfnamefont {P.}~\bibnamefont {{Xie}}}, \bibinfo {author}
  {\bibfnamefont {B.}~\bibnamefont {{Yan}}}, \bibinfo {author} {\bibfnamefont
  {X.}~\bibnamefont {{Yan}}}, \bibinfo {author} {\bibfnamefont
  {J.}~\bibnamefont {{Yang}}}, \bibinfo {author} {\bibfnamefont
  {Y.}~\bibnamefont {{Yang}}}, \bibinfo {author} {\bibfnamefont
  {Y.}~\bibnamefont {{Yao}}}, \bibinfo {author} {\bibfnamefont
  {Z.}~\bibnamefont {{You}}}, \bibinfo {author} {\bibfnamefont
  {C.}~\bibnamefont {{Yu}}}, \bibinfo {author} {\bibfnamefont {J.}~\bibnamefont
  {{Yuan}}}, \bibinfo {author} {\bibfnamefont {Y.}~\bibnamefont {{Yuan}}},
  \bibinfo {author} {\bibfnamefont {Z.}~\bibnamefont {{Yuan}}}, \bibinfo
  {author} {\bibfnamefont {X.}~\bibnamefont {{Zeng}}}, \bibinfo {author}
  {\bibfnamefont {D.}~\bibnamefont {{Zhang}}}, \bibinfo {author} {\bibfnamefont
  {M.}~\bibnamefont {{Zhang}}}, \bibinfo {author} {\bibfnamefont
  {P.}~\bibnamefont {{Zhang}}}, \bibinfo {author} {\bibfnamefont
  {S.}~\bibnamefont {{Zhang}}}, \bibinfo {author} {\bibfnamefont
  {S.}~\bibnamefont {{Zhang}}}, \bibinfo {author} {\bibfnamefont
  {T.}~\bibnamefont {{Zhang}}}, \bibinfo {author} {\bibfnamefont
  {Y.}~\bibnamefont {{Zhang}}}, \bibinfo {author} {\bibfnamefont
  {Y.}~\bibnamefont {{Zhang}}}, \bibinfo {author} {\bibfnamefont
  {Y.}~\bibnamefont {{Zhang}}}, \bibinfo {author} {\bibfnamefont
  {L.}~\bibnamefont {{Zhao}}}, \bibinfo {author} {\bibfnamefont
  {Q.}~\bibnamefont {{Zheng}}}, \bibinfo {author} {\bibfnamefont
  {J.}~\bibnamefont {{Zhou}}}, \bibinfo {author} {\bibfnamefont
  {N.}~\bibnamefont {{Zhou}}}, \bibinfo {author} {\bibfnamefont
  {X.}~\bibnamefont {{Zhou}}}, \bibinfo {author} {\bibfnamefont
  {Y.}~\bibnamefont {{Zhou}}}, \bibinfo {author} {\bibfnamefont
  {Y.}~\bibnamefont {{Zhou}}},\ and\ \bibinfo {author} {\bibnamefont {{PandaX
  Collaboration}}},\ }\bibfield  {title} {\bibinfo {title} {{Search for Light
  Dark Matter with Ionization Signals in the PandaX-4T Experiment}},\ }\href
  {https://doi.org/10.1103/PhysRevLett.130.261001} {\bibfield  {journal}
  {\bibinfo  {journal} {\prl}\ }\textbf {\bibinfo {volume} {130}},\ \bibinfo
  {eid} {261001} (\bibinfo {year} {2023})},\ \Eprint
  {https://arxiv.org/abs/2212.10067} {arXiv:2212.10067 [hep-ex]} \BibitemShut
  {NoStop}%
\bibitem [{\citenamefont {{Fontaine}}\ \emph {et~al.}(2001)\citenamefont
  {{Fontaine}}, \citenamefont {{Brassard}},\ and\ \citenamefont
  {{Bergeron}}}]{Fontaine2001}%
  \BibitemOpen
  \bibfield  {author} {\bibinfo {author} {\bibfnamefont {G.}~\bibnamefont
  {{Fontaine}}}, \bibinfo {author} {\bibfnamefont {P.}~\bibnamefont
  {{Brassard}}},\ and\ \bibinfo {author} {\bibfnamefont {P.}~\bibnamefont
  {{Bergeron}}},\ }\bibfield  {title} {\bibinfo {title} {{The Potential of
  White Dwarf Cosmochronology}},\ }\href {https://doi.org/10.1086/319535}
  {\bibfield  {journal} {\bibinfo  {journal} {\pasp}\ }\textbf {\bibinfo
  {volume} {113}},\ \bibinfo {pages} {409} (\bibinfo {year}
  {2001})}\BibitemShut {NoStop}%
\bibitem [{\citenamefont {{Isern}}\ \emph {et~al.}(2022)\citenamefont
  {{Isern}}, \citenamefont {{Torres}},\ and\ \citenamefont
  {{Rebassa-Mansergas}}}]{Isern2022}%
  \BibitemOpen
  \bibfield  {author} {\bibinfo {author} {\bibfnamefont {J.}~\bibnamefont
  {{Isern}}}, \bibinfo {author} {\bibfnamefont {S.}~\bibnamefont {{Torres}}},\
  and\ \bibinfo {author} {\bibfnamefont {A.}~\bibnamefont
  {{Rebassa-Mansergas}}},\ }\bibfield  {title} {\bibinfo {title} {{White dwarfs
  as Physics laboratories: lights and shadows}},\ }\href
  {https://doi.org/10.3389/fspas.2022.815517} {\bibfield  {journal} {\bibinfo
  {journal} {Frontiers in Astronomy and Space Sciences}\ }\textbf {\bibinfo
  {volume} {9}},\ \bibinfo {eid} {6} (\bibinfo {year} {2022})},\ \Eprint
  {https://arxiv.org/abs/2202.02052} {arXiv:2202.02052 [astro-ph.HE]}
  \BibitemShut {NoStop}%
\bibitem [{\citenamefont {{Winget}}\ and\ \citenamefont
  {{Kepler}}(2008)}]{Winget2008}%
  \BibitemOpen
  \bibfield  {author} {\bibinfo {author} {\bibfnamefont {D.~E.}\ \bibnamefont
  {{Winget}}}\ and\ \bibinfo {author} {\bibfnamefont {S.~O.}\ \bibnamefont
  {{Kepler}}},\ }\bibfield  {title} {\bibinfo {title} {{Pulsating White Dwarf
  Stars and Precision Asteroseismology}},\ }\href
  {https://doi.org/10.1146/annurev.astro.46.060407.145250} {\bibfield
  {journal} {\bibinfo  {journal} {\araa}\ }\textbf {\bibinfo {volume} {46}},\
  \bibinfo {pages} {157} (\bibinfo {year} {2008})},\ \Eprint
  {https://arxiv.org/abs/0806.2573} {arXiv:0806.2573} \BibitemShut {NoStop}%
\bibitem [{\citenamefont {{Fontaine}}\ and\ \citenamefont
  {{Brassard}}(2008)}]{Fontaine2008}%
  \BibitemOpen
  \bibfield  {author} {\bibinfo {author} {\bibfnamefont {G.}~\bibnamefont
  {{Fontaine}}}\ and\ \bibinfo {author} {\bibfnamefont {P.}~\bibnamefont
  {{Brassard}}},\ }\bibfield  {title} {\bibinfo {title} {{The Pulsating White
  Dwarf Stars}},\ }\href {https://doi.org/10.1086/592788} {\bibfield  {journal}
  {\bibinfo  {journal} {\pasp}\ }\textbf {\bibinfo {volume} {120}},\ \bibinfo
  {pages} {1043} (\bibinfo {year} {2008})}\BibitemShut {NoStop}%
\bibitem [{\citenamefont {{Althaus}}\ \emph {et~al.}(2010)\citenamefont
  {{Althaus}}, \citenamefont {{C{\'o}rsico}}, \citenamefont {{Isern}},\ and\
  \citenamefont {{Garc{\'{\i}}a-Berro}}}]{Althaus2010}%
  \BibitemOpen
  \bibfield  {author} {\bibinfo {author} {\bibfnamefont {L.~G.}\ \bibnamefont
  {{Althaus}}}, \bibinfo {author} {\bibfnamefont {A.~H.}\ \bibnamefont
  {{C{\'o}rsico}}}, \bibinfo {author} {\bibfnamefont {J.}~\bibnamefont
  {{Isern}}},\ and\ \bibinfo {author} {\bibfnamefont {E.}~\bibnamefont
  {{Garc{\'{\i}}a-Berro}}},\ }\bibfield  {title} {\bibinfo {title}
  {{Evolutionary and pulsational properties of white dwarf stars}},\ }\href
  {https://doi.org/10.1007/s00159-010-0033-1} {\bibfield  {journal} {\bibinfo
  {journal} {\aapr}\ }\textbf {\bibinfo {volume} {18}},\ \bibinfo {pages} {471}
  (\bibinfo {year} {2010})},\ \Eprint {https://arxiv.org/abs/1007.2659}
  {arXiv:1007.2659 [astro-ph.SR]} \BibitemShut {NoStop}%
\bibitem [{\citenamefont {{Calcaferro}}\ \emph {et~al.}(2017)\citenamefont
  {{Calcaferro}}, \citenamefont {{C{\'o}rsico}},\ and\ \citenamefont
  {{Althaus}}}]{Calcaferro2017}%
  \BibitemOpen
  \bibfield  {author} {\bibinfo {author} {\bibfnamefont {L.~M.}\ \bibnamefont
  {{Calcaferro}}}, \bibinfo {author} {\bibfnamefont {A.~H.}\ \bibnamefont
  {{C{\'o}rsico}}},\ and\ \bibinfo {author} {\bibfnamefont {L.~G.}\
  \bibnamefont {{Althaus}}},\ }\bibfield  {title} {\bibinfo {title} {{Pulsating
  low-mass white dwarfs in the frame of new evolutionary sequences. IV. The
  secular rate of period change}},\ }\href
  {https://doi.org/10.1051/0004-6361/201630376} {\bibfield  {journal} {\bibinfo
   {journal} {\aap}\ }\textbf {\bibinfo {volume} {600}},\ \bibinfo {eid} {A73}
  (\bibinfo {year} {2017})},\ \Eprint {https://arxiv.org/abs/1701.08880}
  {arXiv:1701.08880 [astro-ph.SR]} \BibitemShut {NoStop}%
\bibitem [{\citenamefont {{Winget}}\ \emph {et~al.}(1983)\citenamefont
  {{Winget}}, \citenamefont {{Hansen}},\ and\ \citenamefont {{van
  Horn}}}]{Winget1983}%
  \BibitemOpen
  \bibfield  {author} {\bibinfo {author} {\bibfnamefont {D.~E.}\ \bibnamefont
  {{Winget}}}, \bibinfo {author} {\bibfnamefont {C.~J.}\ \bibnamefont
  {{Hansen}}},\ and\ \bibinfo {author} {\bibfnamefont {H.~M.}\ \bibnamefont
  {{van Horn}}},\ }\bibfield  {title} {\bibinfo {title} {{Do pulsating
  PG1159-035 stars put constraints on stellar evolution?}},\ }\href
  {https://doi.org/10.1038/303781a0} {\bibfield  {journal} {\bibinfo  {journal}
  {\nat}\ }\textbf {\bibinfo {volume} {303}},\ \bibinfo {pages} {781} (\bibinfo
  {year} {1983})}\BibitemShut {NoStop}%
\bibitem [{Note1()}]{Note1}%
  \BibitemOpen
  \bibinfo {note} {Here, we take $P$, $T_{c}$, and $M_{*}$ as
  constants.}\BibitemShut {Stop}%
\bibitem [{\citenamefont {{Niu}}\ \emph
  {et~al.}(2018{\natexlab{b}})\citenamefont {{Niu}}, \citenamefont {{Li}},
  \citenamefont {{Zong}}, \citenamefont {{Xue}},\ and\ \citenamefont
  {{Wang}}}]{Niu2018_dav}%
  \BibitemOpen
  \bibfield  {author} {\bibinfo {author} {\bibfnamefont {J.-S.}\ \bibnamefont
  {{Niu}}}, \bibinfo {author} {\bibfnamefont {T.}~\bibnamefont {{Li}}},
  \bibinfo {author} {\bibfnamefont {W.}~\bibnamefont {{Zong}}}, \bibinfo
  {author} {\bibfnamefont {H.-F.}\ \bibnamefont {{Xue}}},\ and\ \bibinfo
  {author} {\bibfnamefont {Y.}~\bibnamefont {{Wang}}},\ }\bibfield  {title}
  {\bibinfo {title} {{Probing the dark matter-electron interactions via
  hydrogen-atmosphere pulsating white dwarfs}},\ }\href
  {https://doi.org/10.1103/PhysRevD.98.103023} {\bibfield  {journal} {\bibinfo
  {journal} {\prd}\ }\textbf {\bibinfo {volume} {98}},\ \bibinfo {eid} {103023}
  (\bibinfo {year} {2018}{\natexlab{b}})},\ \Eprint
  {https://arxiv.org/abs/1709.08804} {arXiv:1709.08804 [astro-ph.HE]}
  \BibitemShut {NoStop}%
\bibitem [{\citenamefont {{Brickhill}}(1983)}]{Brickhill1983}%
  \BibitemOpen
  \bibfield  {author} {\bibinfo {author} {\bibfnamefont {A.~J.}\ \bibnamefont
  {{Brickhill}}},\ }\bibfield  {title} {\bibinfo {title} {{The pulsations of ZZ
  Ceti stars.}},\ }\href {https://doi.org/10.1093/mnras/204.2.537} {\bibfield
  {journal} {\bibinfo  {journal} {\mnras}\ }\textbf {\bibinfo {volume} {204}},\
  \bibinfo {pages} {537} (\bibinfo {year} {1983})}\BibitemShut {NoStop}%
\bibitem [{\citenamefont {{Mukadam}}\ \emph {et~al.}(2006)\citenamefont
  {{Mukadam}}, \citenamefont {{Montgomery}}, \citenamefont {{Winget}},
  \citenamefont {{Kepler}},\ and\ \citenamefont {{Clemens}}}]{Mukadam2006}%
  \BibitemOpen
  \bibfield  {author} {\bibinfo {author} {\bibfnamefont {A.~S.}\ \bibnamefont
  {{Mukadam}}}, \bibinfo {author} {\bibfnamefont {M.~H.}\ \bibnamefont
  {{Montgomery}}}, \bibinfo {author} {\bibfnamefont {D.~E.}\ \bibnamefont
  {{Winget}}}, \bibinfo {author} {\bibfnamefont {S.~O.}\ \bibnamefont
  {{Kepler}}},\ and\ \bibinfo {author} {\bibfnamefont {J.~C.}\ \bibnamefont
  {{Clemens}}},\ }\bibfield  {title} {\bibinfo {title} {{Ensemble
  Characteristics of the ZZ Ceti Stars}},\ }\href
  {https://doi.org/10.1086/500289} {\bibfield  {journal} {\bibinfo  {journal}
  {\apj}\ }\textbf {\bibinfo {volume} {640}},\ \bibinfo {pages} {956} (\bibinfo
  {year} {2006})},\ \Eprint {https://arxiv.org/abs/astro-ph/0507425}
  {arXiv:astro-ph/0507425 [astro-ph]} \BibitemShut {NoStop}%
\bibitem [{\citenamefont {{C{\'o}rsico}}\ \emph {et~al.}(2016)\citenamefont
  {{C{\'o}rsico}}, \citenamefont {{Romero}}, \citenamefont {{Althaus}},
  \citenamefont {{Garc{\'\i}a-Berro}}, \citenamefont {{Isern}}, \citenamefont
  {{Kepler}}, \citenamefont {{Miller Bertolami}}, \citenamefont {{Sullivan}},\
  and\ \citenamefont {{Chote}}}]{Corsico2016}%
  \BibitemOpen
  \bibfield  {author} {\bibinfo {author} {\bibfnamefont {A.~H.}\ \bibnamefont
  {{C{\'o}rsico}}}, \bibinfo {author} {\bibfnamefont {A.~D.}\ \bibnamefont
  {{Romero}}}, \bibinfo {author} {\bibfnamefont {L.~G.}\ \bibnamefont
  {{Althaus}}}, \bibinfo {author} {\bibfnamefont {E.}~\bibnamefont
  {{Garc{\'\i}a-Berro}}}, \bibinfo {author} {\bibfnamefont {J.}~\bibnamefont
  {{Isern}}}, \bibinfo {author} {\bibfnamefont {S.~O.}\ \bibnamefont
  {{Kepler}}}, \bibinfo {author} {\bibfnamefont {M.~M.}\ \bibnamefont {{Miller
  Bertolami}}}, \bibinfo {author} {\bibfnamefont {D.~J.}\ \bibnamefont
  {{Sullivan}}},\ and\ \bibinfo {author} {\bibfnamefont {P.}~\bibnamefont
  {{Chote}}},\ }\bibfield  {title} {\bibinfo {title} {{An asteroseismic
  constraint on the mass of the axion from the period drift of the pulsating DA
  white dwarf star L19-2}},\ }\href
  {https://doi.org/10.1088/1475-7516/2016/07/036} {\bibfield  {journal}
  {\bibinfo  {journal} {\jcap}\ }\textbf {\bibinfo {volume} {2016}},\ \bibinfo
  {eid} {036} (\bibinfo {year} {2016})},\ \Eprint
  {https://arxiv.org/abs/1605.06458} {arXiv:1605.06458 [astro-ph.SR]}
  \BibitemShut {NoStop}%
\bibitem [{\citenamefont {{Romero}}\ \emph {et~al.}(2012)\citenamefont
  {{Romero}}, \citenamefont {{C{\'o}rsico}}, \citenamefont {{Althaus}},
  \citenamefont {{Kepler}}, \citenamefont {{Castanheira}},\ and\ \citenamefont
  {{Miller Bertolami}}}]{Romero2012}%
  \BibitemOpen
  \bibfield  {author} {\bibinfo {author} {\bibfnamefont {A.~D.}\ \bibnamefont
  {{Romero}}}, \bibinfo {author} {\bibfnamefont {A.~H.}\ \bibnamefont
  {{C{\'o}rsico}}}, \bibinfo {author} {\bibfnamefont {L.~G.}\ \bibnamefont
  {{Althaus}}}, \bibinfo {author} {\bibfnamefont {S.~O.}\ \bibnamefont
  {{Kepler}}}, \bibinfo {author} {\bibfnamefont {B.~G.}\ \bibnamefont
  {{Castanheira}}},\ and\ \bibinfo {author} {\bibfnamefont {M.~M.}\
  \bibnamefont {{Miller Bertolami}}},\ }\bibfield  {title} {\bibinfo {title}
  {{Toward ensemble asteroseismology of ZZ Ceti stars with fully evolutionary
  models}},\ }\href {https://doi.org/10.1111/j.1365-2966.2011.20134.x}
  {\bibfield  {journal} {\bibinfo  {journal} {\mnras}\ }\textbf {\bibinfo
  {volume} {420}},\ \bibinfo {pages} {1462} (\bibinfo {year} {2012})},\ \Eprint
  {https://arxiv.org/abs/1109.6682} {arXiv:1109.6682 [astro-ph.SR]}
  \BibitemShut {NoStop}%
\bibitem [{\citenamefont {{Sullivan}}\ and\ \citenamefont
  {{Chote}}(2015)}]{Sullivan2015}%
  \BibitemOpen
  \bibfield  {author} {\bibinfo {author} {\bibfnamefont {D.~J.}\ \bibnamefont
  {{Sullivan}}}\ and\ \bibinfo {author} {\bibfnamefont {P.}~\bibnamefont
  {{Chote}}},\ }\bibfield  {title} {\bibinfo {title} {{The Frequency Stability
  of the Pulsating White Dwarf L19-2}},\ }in\ \href@noop {} {\emph {\bibinfo
  {booktitle} {19th European Workshop on White Dwarfs}}},\ \bibinfo {series}
  {Astronomical Society of the Pacific Conference Series}, Vol.\ \bibinfo
  {volume} {493},\ \bibinfo {editor} {edited by\ \bibinfo {editor}
  {\bibfnamefont {P.}~\bibnamefont {{Dufour}}}, \bibinfo {editor}
  {\bibfnamefont {P.}~\bibnamefont {{Bergeron}}},\ and\ \bibinfo {editor}
  {\bibfnamefont {G.}~\bibnamefont {{Fontaine}}}}\ (\bibinfo {year} {2015})\
  p.\ \bibinfo {pages} {199}\BibitemShut {NoStop}%
\bibitem [{\citenamefont {{Kepler}}\ \emph {et~al.}(2021)\citenamefont
  {{Kepler}}, \citenamefont {{Winget}}, \citenamefont {{Vanderbosch}},
  \citenamefont {{Castanheira}}, \citenamefont {{Hermes}}, \citenamefont
  {{Bell}}, \citenamefont {{Mullally}}, \citenamefont {{Romero}}, \citenamefont
  {{Montgomery}}, \citenamefont {{DeGennaro}}, \citenamefont {{Winget}},
  \citenamefont {{Chandler}}, \citenamefont {{Jeffery}}, \citenamefont
  {{Fritzen}}, \citenamefont {{Williams}}, \citenamefont {{Chote}},\ and\
  \citenamefont {{Zola}}}]{Kepler2021}%
  \BibitemOpen
  \bibfield  {author} {\bibinfo {author} {\bibfnamefont {S.~O.}\ \bibnamefont
  {{Kepler}}}, \bibinfo {author} {\bibfnamefont {D.~E.}\ \bibnamefont
  {{Winget}}}, \bibinfo {author} {\bibfnamefont {Z.~P.}\ \bibnamefont
  {{Vanderbosch}}}, \bibinfo {author} {\bibfnamefont {B.~G.}\ \bibnamefont
  {{Castanheira}}}, \bibinfo {author} {\bibfnamefont {J.~J.}\ \bibnamefont
  {{Hermes}}}, \bibinfo {author} {\bibfnamefont {K.~J.}\ \bibnamefont
  {{Bell}}}, \bibinfo {author} {\bibfnamefont {F.}~\bibnamefont {{Mullally}}},
  \bibinfo {author} {\bibfnamefont {A.~D.}\ \bibnamefont {{Romero}}}, \bibinfo
  {author} {\bibfnamefont {M.~H.}\ \bibnamefont {{Montgomery}}}, \bibinfo
  {author} {\bibfnamefont {S.}~\bibnamefont {{DeGennaro}}}, \bibinfo {author}
  {\bibfnamefont {K.~I.}\ \bibnamefont {{Winget}}}, \bibinfo {author}
  {\bibfnamefont {D.}~\bibnamefont {{Chandler}}}, \bibinfo {author}
  {\bibfnamefont {E.~J.}\ \bibnamefont {{Jeffery}}}, \bibinfo {author}
  {\bibfnamefont {J.~K.}\ \bibnamefont {{Fritzen}}}, \bibinfo {author}
  {\bibfnamefont {K.~A.}\ \bibnamefont {{Williams}}}, \bibinfo {author}
  {\bibfnamefont {P.}~\bibnamefont {{Chote}}},\ and\ \bibinfo {author}
  {\bibfnamefont {S.}~\bibnamefont {{Zola}}},\ }\bibfield  {title} {\bibinfo
  {title} {{The Pulsating White Dwarf G117-B15A: Still the Most Stable Optical
  Clock Known}},\ }\href {https://doi.org/10.3847/1538-4357/abc626} {\bibfield
  {journal} {\bibinfo  {journal} {\apj}\ }\textbf {\bibinfo {volume} {906}},\
  \bibinfo {eid} {7} (\bibinfo {year} {2021})},\ \Eprint
  {https://arxiv.org/abs/2010.16062} {arXiv:2010.16062 [astro-ph.SR]}
  \BibitemShut {NoStop}%
\bibitem [{\citenamefont {{Bailer-Jones}}\ \emph {et~al.}(2021)\citenamefont
  {{Bailer-Jones}}, \citenamefont {{Rybizki}}, \citenamefont {{Fouesneau}},
  \citenamefont {{Demleitner}},\ and\ \citenamefont {{Andrae}}}]{Bailer2021}%
  \BibitemOpen
  \bibfield  {author} {\bibinfo {author} {\bibfnamefont {C.~A.~L.}\
  \bibnamefont {{Bailer-Jones}}}, \bibinfo {author} {\bibfnamefont
  {J.}~\bibnamefont {{Rybizki}}}, \bibinfo {author} {\bibfnamefont
  {M.}~\bibnamefont {{Fouesneau}}}, \bibinfo {author} {\bibfnamefont
  {M.}~\bibnamefont {{Demleitner}}},\ and\ \bibinfo {author} {\bibfnamefont
  {R.}~\bibnamefont {{Andrae}}},\ }\bibfield  {title} {\bibinfo {title}
  {{Estimating Distances from Parallaxes. V. Geometric and Photogeometric
  Distances to 1.47 Billion Stars in Gaia Early Data Release 3}},\ }\href
  {https://doi.org/10.3847/1538-3881/abd806} {\bibfield  {journal} {\bibinfo
  {journal} {\aj}\ }\textbf {\bibinfo {volume} {161}},\ \bibinfo {eid} {147}
  (\bibinfo {year} {2021})},\ \Eprint {https://arxiv.org/abs/2012.05220}
  {arXiv:2012.05220 [astro-ph.SR]} \BibitemShut {NoStop}%
\bibitem [{\citenamefont {{Mukadam}}\ \emph {et~al.}(2013)\citenamefont
  {{Mukadam}}, \citenamefont {{Bischoff-Kim}}, \citenamefont {{Fraser}},
  \citenamefont {{C{\'o}rsico}}, \citenamefont {{Montgomery}}, \citenamefont
  {{Kepler}}, \citenamefont {{Romero}}, \citenamefont {{Winget}}, \citenamefont
  {{Hermes}}, \citenamefont {{Riecken}}, \citenamefont {{Kronberg}},
  \citenamefont {{Winget}}, \citenamefont {{Falcon}}, \citenamefont
  {{Chandler}}, \citenamefont {{Kuehne}}, \citenamefont {{Sullivan}},
  \citenamefont {{Reaves}}, \citenamefont {{von Hippel}}, \citenamefont
  {{Mullally}}, \citenamefont {{Shipman}}, \citenamefont {{Thompson}},
  \citenamefont {{Silvestri}},\ and\ \citenamefont {{Hynes}}}]{Mukadam2013}%
  \BibitemOpen
  \bibfield  {author} {\bibinfo {author} {\bibfnamefont {A.~S.}\ \bibnamefont
  {{Mukadam}}}, \bibinfo {author} {\bibfnamefont {A.}~\bibnamefont
  {{Bischoff-Kim}}}, \bibinfo {author} {\bibfnamefont {O.}~\bibnamefont
  {{Fraser}}}, \bibinfo {author} {\bibfnamefont {A.~H.}\ \bibnamefont
  {{C{\'o}rsico}}}, \bibinfo {author} {\bibfnamefont {M.~H.}\ \bibnamefont
  {{Montgomery}}}, \bibinfo {author} {\bibfnamefont {S.~O.}\ \bibnamefont
  {{Kepler}}}, \bibinfo {author} {\bibfnamefont {A.~D.}\ \bibnamefont
  {{Romero}}}, \bibinfo {author} {\bibfnamefont {D.~E.}\ \bibnamefont
  {{Winget}}}, \bibinfo {author} {\bibfnamefont {J.~J.}\ \bibnamefont
  {{Hermes}}}, \bibinfo {author} {\bibfnamefont {T.~S.}\ \bibnamefont
  {{Riecken}}}, \bibinfo {author} {\bibfnamefont {M.~E.}\ \bibnamefont
  {{Kronberg}}}, \bibinfo {author} {\bibfnamefont {K.~I.}\ \bibnamefont
  {{Winget}}}, \bibinfo {author} {\bibfnamefont {R.~E.}\ \bibnamefont
  {{Falcon}}}, \bibinfo {author} {\bibfnamefont {D.~W.}\ \bibnamefont
  {{Chandler}}}, \bibinfo {author} {\bibfnamefont {J.~W.}\ \bibnamefont
  {{Kuehne}}}, \bibinfo {author} {\bibfnamefont {D.~J.}\ \bibnamefont
  {{Sullivan}}}, \bibinfo {author} {\bibfnamefont {D.}~\bibnamefont
  {{Reaves}}}, \bibinfo {author} {\bibfnamefont {T.}~\bibnamefont {{von
  Hippel}}}, \bibinfo {author} {\bibfnamefont {F.}~\bibnamefont {{Mullally}}},
  \bibinfo {author} {\bibfnamefont {H.}~\bibnamefont {{Shipman}}}, \bibinfo
  {author} {\bibfnamefont {S.~E.}\ \bibnamefont {{Thompson}}}, \bibinfo
  {author} {\bibfnamefont {N.~M.}\ \bibnamefont {{Silvestri}}},\ and\ \bibinfo
  {author} {\bibfnamefont {R.~I.}\ \bibnamefont {{Hynes}}},\ }\bibfield
  {title} {\bibinfo {title} {{Measuring the Evolutionary Rate of Cooling of ZZ
  Ceti}},\ }\href {https://doi.org/10.1088/0004-637X/771/1/17} {\bibfield
  {journal} {\bibinfo  {journal} {\apj}\ }\textbf {\bibinfo {volume} {771}},\
  \bibinfo {eid} {17} (\bibinfo {year} {2013})}\BibitemShut {NoStop}%
\bibitem [{\citenamefont {{Pajdosz}}(1995)}]{Pajdosz1995}%
  \BibitemOpen
  \bibfield  {author} {\bibinfo {author} {\bibfnamefont {G.}~\bibnamefont
  {{Pajdosz}}},\ }\bibfield  {title} {\bibinfo {title} {{Non-evolutionary
  secular period increase in pulsating DA white dwarfs.}},\ }\href@noop {}
  {\bibfield  {journal} {\bibinfo  {journal} {\aap}\ }\textbf {\bibinfo
  {volume} {295}},\ \bibinfo {pages} {L17} (\bibinfo {year}
  {1995})}\BibitemShut {NoStop}%
\bibitem [{\citenamefont {{Abe}}\ \emph {et~al.}(2023)\citenamefont {{Abe}},
  \citenamefont {{Hayato}}, \citenamefont {{Hiraide}}, \citenamefont {{Ieki}},
  \citenamefont {{Ikeda}}, \citenamefont {{Kameda}}, \citenamefont
  {{Kanemura}}, \citenamefont {{Kaneshima}}, \citenamefont {{Kashiwagi}},
  \citenamefont {{Kataoka}}, \citenamefont {{Miki}}, \citenamefont {{Mine}},
  \citenamefont {{Miura}}, \citenamefont {{Moriyama}}, \citenamefont
  {{Nakano}}, \citenamefont {{Nakahata}}, \citenamefont {{Nakayama}},
  \citenamefont {{Noguchi}}, \citenamefont {{Okamoto}}, \citenamefont {{Sato}},
  \citenamefont {{Sekiya}}, \citenamefont {{Shiba}}, \citenamefont {{Shimizu}},
  \citenamefont {{Shiozawa}}, \citenamefont {{Sonoda}}, \citenamefont
  {{Suzuki}}, \citenamefont {{Takeda}}, \citenamefont {{Takemoto}},
  \citenamefont {{Takenaka}}, \citenamefont {{Tanaka}}, \citenamefont
  {{Watanabe}}, \citenamefont {{Yano}}, \citenamefont {{Han}}, \citenamefont
  {{Kajita}}, \citenamefont {{Okumura}}, \citenamefont {{Tashiro}},
  \citenamefont {{Tomiya}}, \citenamefont {{Wang}}, \citenamefont {{Xia}},
  \citenamefont {{Yoshida}}, \citenamefont {{Megias}}, \citenamefont
  {{Fernandez}}, \citenamefont {{Labarga}}, \citenamefont {{Ospina}},
  \citenamefont {{Zaldivar}}, \citenamefont {{Pointon}}, \citenamefont
  {{Kearns}}, \citenamefont {{Raaf}}, \citenamefont {{Wan}}, \citenamefont
  {{Wester}}, \citenamefont {{Bian}}, \citenamefont {{Griskevich}},
  \citenamefont {{Kropp}}, \citenamefont {{Locke}}, \citenamefont {{Smy}},
  \citenamefont {{Sobel}}, \citenamefont {{Takhistov}}, \citenamefont
  {{Yankelevich}}, \citenamefont {{Hill}}, \citenamefont {{Park}},
  \citenamefont {{Bodur}}, \citenamefont {{Scholberg}}, \citenamefont
  {{Walter}}, \citenamefont {{Bernard}}, \citenamefont {{Coffani}},
  \citenamefont {{Drapier}}, \citenamefont {{El Hedri}}, \citenamefont
  {{Giampaolo}}, \citenamefont {{Mueller}}, \citenamefont {{Santos}},
  \citenamefont {{Paganini}}, \citenamefont {{Quilain}}, \citenamefont
  {{Ishizuka}}, \citenamefont {{Nakamura}}, \citenamefont {{Jang}},
  \citenamefont {{Learned}}, \citenamefont {{Choi}}, \citenamefont {{Cao}},
  \citenamefont {{Anthony}}, \citenamefont {{Martin}}, \citenamefont {{Scott}},
  \citenamefont {{Sztuc}}, \citenamefont {{Uchida}}, \citenamefont {{Berardi}},
  \citenamefont {{Catanesi}}, \citenamefont {{Radicioni}}, \citenamefont
  {{Calabria}}, \citenamefont {{Machado}}, \citenamefont {{De Rosa}},
  \citenamefont {{Collazuol}}, \citenamefont {{Iacob}}, \citenamefont
  {{Lamoureux}}, \citenamefont {{Mattiazzi}}, \citenamefont {{Ludovici}},
  \citenamefont {{Gonin}}, \citenamefont {{Pronost}}, \citenamefont
  {{Fujisawa}}, \citenamefont {{Maekawa}}, \citenamefont {{Nishimura}},
  \citenamefont {{Friend}}, \citenamefont {{Hasegawa}}, \citenamefont
  {{Ishida}}, \citenamefont {{Kobayashi}}, \citenamefont {{Jakkapu}},
  \citenamefont {{Matsubara}}, \citenamefont {{Nakadaira}}, \citenamefont
  {{Nakamura}}, \citenamefont {{Oyama}}, \citenamefont {{Sakashita}},
  \citenamefont {{Sekiguchi}}, \citenamefont {{Tsukamoto}}, \citenamefont
  {{Boschi}}, \citenamefont {{Di Lodovico}}, \citenamefont {{Gao}},
  \citenamefont {{Goldsack}}, \citenamefont {{Katori}}, \citenamefont
  {{Migenda}}, \citenamefont {{Taani}}, \citenamefont {{Zsoldos}},
  \citenamefont {{Kotsar}}, \citenamefont {{Ozaki}}, \citenamefont {{Suzuki}},
  \citenamefont {{Takeuchi}}, \citenamefont {{Bronner}}, \citenamefont
  {{Feng}}, \citenamefont {{Kikawa}}, \citenamefont {{Mori}}, \citenamefont
  {{Nakaya}}, \citenamefont {{Wendell}}, \citenamefont {{Yasutome}},
  \citenamefont {{Jenkins}}, \citenamefont {{McCauley}}, \citenamefont
  {{Mehta}}, \citenamefont {{Tsui}}, \citenamefont {{Fukuda}}, \citenamefont
  {{Itow}}, \citenamefont {{Menjo}}, \citenamefont {{Ninomiya}}, \citenamefont
  {{Lagoda}}, \citenamefont {{Lakshmi}}, \citenamefont {{Mandal}},
  \citenamefont {{Mijakowski}}, \citenamefont {{Prabhu}}, \citenamefont
  {{Zalipska}}, \citenamefont {{Jia}}, \citenamefont {{Jiang}}, \citenamefont
  {{Jung}}, \citenamefont {{Wilking}}, \citenamefont {{Yanagisawa}},
  \citenamefont {{Harada}}, \citenamefont {{Ishino}}, \citenamefont {{Ito}},
  \citenamefont {{Kitagawa}}, \citenamefont {{Koshio}}, \citenamefont
  {{Nakanishi}}, \citenamefont {{Sakai}}, \citenamefont {{Barr}}, \citenamefont
  {{Barrow}}, \citenamefont {{Cook}}, \citenamefont {{Samani}}, \citenamefont
  {{Wark}}, \citenamefont {{Nova}}, \citenamefont {{Yang}}, \citenamefont
  {{Malek}}, \citenamefont {{McElwee}}, \citenamefont {{Stone}}, \citenamefont
  {{Thiesse}}, \citenamefont {{Thompson}}, \citenamefont {{Okazawa}},
  \citenamefont {{Kim}}, \citenamefont {{Seo}}, \citenamefont {{Yu}},
  \citenamefont {{Ichikawa}}, \citenamefont {{Nakamura}}, \citenamefont
  {{Tairafune}}, \citenamefont {{Nishijima}}, \citenamefont {{Iwamoto}},
  \citenamefont {{Nakagiri}}, \citenamefont {{Nakajima}}, \citenamefont
  {{Taniuchi}}, \citenamefont {{Yokoyama}}, \citenamefont {{Martens}},
  \citenamefont {{de Perio}}, \citenamefont {{Vagins}}, \citenamefont {{Kuze}},
  \citenamefont {{Izumiyama}}, \citenamefont {{Inomoto}}, \citenamefont
  {{Ishitsuka}}, \citenamefont {{Ito}}, \citenamefont {{Kinoshita}},
  \citenamefont {{Matsumoto}}, \citenamefont {{Ommura}}, \citenamefont
  {{Shigeta}}, \citenamefont {{Shinoki}}, \citenamefont {{Suganuma}},
  \citenamefont {{Yamauchi}}, \citenamefont {{Martin}}, \citenamefont
  {{Tanaka}}, \citenamefont {{Towstego}}, \citenamefont {{Akutsu}},
  \citenamefont {{Gousy-Leblanc}}, \citenamefont {{Hartz}}, \citenamefont
  {{Konaka}}, \citenamefont {{Prouse}}, \citenamefont {{Chen}}, \citenamefont
  {{Xu}}, \citenamefont {{Zhang}}, \citenamefont {{Posiadala-Zezula}},
  \citenamefont {{Hadley}}, \citenamefont {{Nicholson}}, \citenamefont
  {{O'Flaherty}}, \citenamefont {{Richards}}, \citenamefont {{Ali}},
  \citenamefont {{Jamieson}}, \citenamefont {{Marti}}, \citenamefont
  {{Minamino}}, \citenamefont {{Pintaudi}}, \citenamefont {{Sano}},
  \citenamefont {{Suzuki}}, \citenamefont {{Wada}},\ and\ \citenamefont
  {{Super-Kamiokande Collaboration}}}]{DMMeV2023}%
  \BibitemOpen
  \bibfield  {author} {\bibinfo {author} {\bibfnamefont {K.}~\bibnamefont
  {{Abe}}}, \bibinfo {author} {\bibfnamefont {Y.}~\bibnamefont {{Hayato}}},
  \bibinfo {author} {\bibfnamefont {K.}~\bibnamefont {{Hiraide}}}, \bibinfo
  {author} {\bibfnamefont {K.}~\bibnamefont {{Ieki}}}, \bibinfo {author}
  {\bibfnamefont {M.}~\bibnamefont {{Ikeda}}}, \bibinfo {author} {\bibfnamefont
  {J.}~\bibnamefont {{Kameda}}}, \bibinfo {author} {\bibfnamefont
  {Y.}~\bibnamefont {{Kanemura}}}, \bibinfo {author} {\bibfnamefont
  {R.}~\bibnamefont {{Kaneshima}}}, \bibinfo {author} {\bibfnamefont
  {Y.}~\bibnamefont {{Kashiwagi}}}, \bibinfo {author} {\bibfnamefont
  {Y.}~\bibnamefont {{Kataoka}}}, \bibinfo {author} {\bibfnamefont
  {S.}~\bibnamefont {{Miki}}}, \bibinfo {author} {\bibfnamefont
  {S.}~\bibnamefont {{Mine}}}, \bibinfo {author} {\bibfnamefont
  {M.}~\bibnamefont {{Miura}}}, \bibinfo {author} {\bibfnamefont
  {S.}~\bibnamefont {{Moriyama}}}, \bibinfo {author} {\bibfnamefont
  {Y.}~\bibnamefont {{Nakano}}}, \bibinfo {author} {\bibfnamefont
  {M.}~\bibnamefont {{Nakahata}}}, \bibinfo {author} {\bibfnamefont
  {S.}~\bibnamefont {{Nakayama}}}, \bibinfo {author} {\bibfnamefont
  {Y.}~\bibnamefont {{Noguchi}}}, \bibinfo {author} {\bibfnamefont
  {K.}~\bibnamefont {{Okamoto}}}, \bibinfo {author} {\bibfnamefont
  {K.}~\bibnamefont {{Sato}}}, \bibinfo {author} {\bibfnamefont
  {H.}~\bibnamefont {{Sekiya}}}, \bibinfo {author} {\bibfnamefont
  {H.}~\bibnamefont {{Shiba}}}, \bibinfo {author} {\bibfnamefont
  {K.}~\bibnamefont {{Shimizu}}}, \bibinfo {author} {\bibfnamefont
  {M.}~\bibnamefont {{Shiozawa}}}, \bibinfo {author} {\bibfnamefont
  {Y.}~\bibnamefont {{Sonoda}}}, \bibinfo {author} {\bibfnamefont
  {Y.}~\bibnamefont {{Suzuki}}}, \bibinfo {author} {\bibfnamefont
  {A.}~\bibnamefont {{Takeda}}}, \bibinfo {author} {\bibfnamefont
  {Y.}~\bibnamefont {{Takemoto}}}, \bibinfo {author} {\bibfnamefont
  {A.}~\bibnamefont {{Takenaka}}}, \bibinfo {author} {\bibfnamefont
  {H.}~\bibnamefont {{Tanaka}}}, \bibinfo {author} {\bibfnamefont
  {S.}~\bibnamefont {{Watanabe}}}, \bibinfo {author} {\bibfnamefont
  {T.}~\bibnamefont {{Yano}}}, \bibinfo {author} {\bibfnamefont
  {S.}~\bibnamefont {{Han}}}, \bibinfo {author} {\bibfnamefont
  {T.}~\bibnamefont {{Kajita}}}, \bibinfo {author} {\bibfnamefont
  {K.}~\bibnamefont {{Okumura}}}, \bibinfo {author} {\bibfnamefont
  {T.}~\bibnamefont {{Tashiro}}}, \bibinfo {author} {\bibfnamefont
  {T.}~\bibnamefont {{Tomiya}}}, \bibinfo {author} {\bibfnamefont
  {X.}~\bibnamefont {{Wang}}}, \bibinfo {author} {\bibfnamefont
  {J.}~\bibnamefont {{Xia}}}, \bibinfo {author} {\bibfnamefont
  {S.}~\bibnamefont {{Yoshida}}}, \bibinfo {author} {\bibfnamefont {G.~D.}\
  \bibnamefont {{Megias}}}, \bibinfo {author} {\bibfnamefont {P.}~\bibnamefont
  {{Fernandez}}}, \bibinfo {author} {\bibfnamefont {L.}~\bibnamefont
  {{Labarga}}}, \bibinfo {author} {\bibfnamefont {N.}~\bibnamefont {{Ospina}}},
  \bibinfo {author} {\bibfnamefont {B.}~\bibnamefont {{Zaldivar}}}, \bibinfo
  {author} {\bibfnamefont {B.~W.}\ \bibnamefont {{Pointon}}}, \bibinfo {author}
  {\bibfnamefont {E.}~\bibnamefont {{Kearns}}}, \bibinfo {author}
  {\bibfnamefont {J.~L.}\ \bibnamefont {{Raaf}}}, \bibinfo {author}
  {\bibfnamefont {L.}~\bibnamefont {{Wan}}}, \bibinfo {author} {\bibfnamefont
  {T.}~\bibnamefont {{Wester}}}, \bibinfo {author} {\bibfnamefont
  {J.}~\bibnamefont {{Bian}}}, \bibinfo {author} {\bibfnamefont {N.~J.}\
  \bibnamefont {{Griskevich}}}, \bibinfo {author} {\bibfnamefont {W.~R.}\
  \bibnamefont {{Kropp}}}, \bibinfo {author} {\bibfnamefont {S.}~\bibnamefont
  {{Locke}}}, \bibinfo {author} {\bibfnamefont {M.~B.}\ \bibnamefont {{Smy}}},
  \bibinfo {author} {\bibfnamefont {H.~W.}\ \bibnamefont {{Sobel}}}, \bibinfo
  {author} {\bibfnamefont {V.}~\bibnamefont {{Takhistov}}}, \bibinfo {author}
  {\bibfnamefont {A.}~\bibnamefont {{Yankelevich}}}, \bibinfo {author}
  {\bibfnamefont {J.}~\bibnamefont {{Hill}}}, \bibinfo {author} {\bibfnamefont
  {R.~G.}\ \bibnamefont {{Park}}}, \bibinfo {author} {\bibfnamefont
  {B.}~\bibnamefont {{Bodur}}}, \bibinfo {author} {\bibfnamefont
  {K.}~\bibnamefont {{Scholberg}}}, \bibinfo {author} {\bibfnamefont {C.~W.}\
  \bibnamefont {{Walter}}}, \bibinfo {author} {\bibfnamefont {L.}~\bibnamefont
  {{Bernard}}}, \bibinfo {author} {\bibfnamefont {A.}~\bibnamefont
  {{Coffani}}}, \bibinfo {author} {\bibfnamefont {O.}~\bibnamefont
  {{Drapier}}}, \bibinfo {author} {\bibfnamefont {S.}~\bibnamefont {{El
  Hedri}}}, \bibinfo {author} {\bibfnamefont {A.}~\bibnamefont {{Giampaolo}}},
  \bibinfo {author} {\bibfnamefont {T.~A.}\ \bibnamefont {{Mueller}}}, \bibinfo
  {author} {\bibfnamefont {A.~D.}\ \bibnamefont {{Santos}}}, \bibinfo {author}
  {\bibfnamefont {P.}~\bibnamefont {{Paganini}}}, \bibinfo {author}
  {\bibfnamefont {B.}~\bibnamefont {{Quilain}}}, \bibinfo {author}
  {\bibfnamefont {T.}~\bibnamefont {{Ishizuka}}}, \bibinfo {author}
  {\bibfnamefont {T.}~\bibnamefont {{Nakamura}}}, \bibinfo {author}
  {\bibfnamefont {J.~S.}\ \bibnamefont {{Jang}}}, \bibinfo {author}
  {\bibfnamefont {J.~G.}\ \bibnamefont {{Learned}}}, \bibinfo {author}
  {\bibfnamefont {K.}~\bibnamefont {{Choi}}}, \bibinfo {author} {\bibfnamefont
  {S.}~\bibnamefont {{Cao}}}, \bibinfo {author} {\bibfnamefont {L.~H.~V.}\
  \bibnamefont {{Anthony}}}, \bibinfo {author} {\bibfnamefont {D.}~\bibnamefont
  {{Martin}}}, \bibinfo {author} {\bibfnamefont {M.}~\bibnamefont {{Scott}}},
  \bibinfo {author} {\bibfnamefont {A.~A.}\ \bibnamefont {{Sztuc}}}, \bibinfo
  {author} {\bibfnamefont {Y.}~\bibnamefont {{Uchida}}}, \bibinfo {author}
  {\bibfnamefont {V.}~\bibnamefont {{Berardi}}}, \bibinfo {author}
  {\bibfnamefont {M.~G.}\ \bibnamefont {{Catanesi}}}, \bibinfo {author}
  {\bibfnamefont {E.}~\bibnamefont {{Radicioni}}}, \bibinfo {author}
  {\bibfnamefont {N.~F.}\ \bibnamefont {{Calabria}}}, \bibinfo {author}
  {\bibfnamefont {L.~N.}\ \bibnamefont {{Machado}}}, \bibinfo {author}
  {\bibfnamefont {G.}~\bibnamefont {{De Rosa}}}, \bibinfo {author}
  {\bibfnamefont {G.}~\bibnamefont {{Collazuol}}}, \bibinfo {author}
  {\bibfnamefont {F.}~\bibnamefont {{Iacob}}}, \bibinfo {author} {\bibfnamefont
  {M.}~\bibnamefont {{Lamoureux}}}, \bibinfo {author} {\bibfnamefont
  {M.}~\bibnamefont {{Mattiazzi}}}, \bibinfo {author} {\bibfnamefont
  {L.}~\bibnamefont {{Ludovici}}}, \bibinfo {author} {\bibfnamefont
  {M.}~\bibnamefont {{Gonin}}}, \bibinfo {author} {\bibfnamefont
  {G.}~\bibnamefont {{Pronost}}}, \bibinfo {author} {\bibfnamefont
  {C.}~\bibnamefont {{Fujisawa}}}, \bibinfo {author} {\bibfnamefont
  {Y.}~\bibnamefont {{Maekawa}}}, \bibinfo {author} {\bibfnamefont
  {Y.}~\bibnamefont {{Nishimura}}}, \bibinfo {author} {\bibfnamefont
  {M.}~\bibnamefont {{Friend}}}, \bibinfo {author} {\bibfnamefont
  {T.}~\bibnamefont {{Hasegawa}}}, \bibinfo {author} {\bibfnamefont
  {T.}~\bibnamefont {{Ishida}}}, \bibinfo {author} {\bibfnamefont
  {T.}~\bibnamefont {{Kobayashi}}}, \bibinfo {author} {\bibfnamefont
  {M.}~\bibnamefont {{Jakkapu}}}, \bibinfo {author} {\bibfnamefont
  {T.}~\bibnamefont {{Matsubara}}}, \bibinfo {author} {\bibfnamefont
  {T.}~\bibnamefont {{Nakadaira}}}, \bibinfo {author} {\bibfnamefont
  {K.}~\bibnamefont {{Nakamura}}}, \bibinfo {author} {\bibfnamefont
  {Y.}~\bibnamefont {{Oyama}}}, \bibinfo {author} {\bibfnamefont
  {K.}~\bibnamefont {{Sakashita}}}, \bibinfo {author} {\bibfnamefont
  {T.}~\bibnamefont {{Sekiguchi}}}, \bibinfo {author} {\bibfnamefont
  {T.}~\bibnamefont {{Tsukamoto}}}, \bibinfo {author} {\bibfnamefont
  {T.}~\bibnamefont {{Boschi}}}, \bibinfo {author} {\bibfnamefont
  {F.}~\bibnamefont {{Di Lodovico}}}, \bibinfo {author} {\bibfnamefont
  {J.}~\bibnamefont {{Gao}}}, \bibinfo {author} {\bibfnamefont
  {A.}~\bibnamefont {{Goldsack}}}, \bibinfo {author} {\bibfnamefont
  {T.}~\bibnamefont {{Katori}}}, \bibinfo {author} {\bibfnamefont
  {J.}~\bibnamefont {{Migenda}}}, \bibinfo {author} {\bibfnamefont
  {M.}~\bibnamefont {{Taani}}}, \bibinfo {author} {\bibfnamefont
  {S.}~\bibnamefont {{Zsoldos}}}, \bibinfo {author} {\bibfnamefont
  {Y.}~\bibnamefont {{Kotsar}}}, \bibinfo {author} {\bibfnamefont
  {H.}~\bibnamefont {{Ozaki}}}, \bibinfo {author} {\bibfnamefont {A.~T.}\
  \bibnamefont {{Suzuki}}}, \bibinfo {author} {\bibfnamefont {Y.}~\bibnamefont
  {{Takeuchi}}}, \bibinfo {author} {\bibfnamefont {C.}~\bibnamefont
  {{Bronner}}}, \bibinfo {author} {\bibfnamefont {J.}~\bibnamefont {{Feng}}},
  \bibinfo {author} {\bibfnamefont {T.}~\bibnamefont {{Kikawa}}}, \bibinfo
  {author} {\bibfnamefont {M.}~\bibnamefont {{Mori}}}, \bibinfo {author}
  {\bibfnamefont {T.}~\bibnamefont {{Nakaya}}}, \bibinfo {author}
  {\bibfnamefont {R.~A.}\ \bibnamefont {{Wendell}}}, \bibinfo {author}
  {\bibfnamefont {K.}~\bibnamefont {{Yasutome}}}, \bibinfo {author}
  {\bibfnamefont {S.~J.}\ \bibnamefont {{Jenkins}}}, \bibinfo {author}
  {\bibfnamefont {N.}~\bibnamefont {{McCauley}}}, \bibinfo {author}
  {\bibfnamefont {P.}~\bibnamefont {{Mehta}}}, \bibinfo {author} {\bibfnamefont
  {K.~M.}\ \bibnamefont {{Tsui}}}, \bibinfo {author} {\bibfnamefont
  {Y.}~\bibnamefont {{Fukuda}}}, \bibinfo {author} {\bibfnamefont
  {Y.}~\bibnamefont {{Itow}}}, \bibinfo {author} {\bibfnamefont
  {H.}~\bibnamefont {{Menjo}}}, \bibinfo {author} {\bibfnamefont
  {K.}~\bibnamefont {{Ninomiya}}}, \bibinfo {author} {\bibfnamefont
  {J.}~\bibnamefont {{Lagoda}}}, \bibinfo {author} {\bibfnamefont {S.~M.}\
  \bibnamefont {{Lakshmi}}}, \bibinfo {author} {\bibfnamefont {M.}~\bibnamefont
  {{Mandal}}}, \bibinfo {author} {\bibfnamefont {P.}~\bibnamefont
  {{Mijakowski}}}, \bibinfo {author} {\bibfnamefont {Y.~S.}\ \bibnamefont
  {{Prabhu}}}, \bibinfo {author} {\bibfnamefont {J.}~\bibnamefont
  {{Zalipska}}}, \bibinfo {author} {\bibfnamefont {M.}~\bibnamefont {{Jia}}},
  \bibinfo {author} {\bibfnamefont {J.}~\bibnamefont {{Jiang}}}, \bibinfo
  {author} {\bibfnamefont {C.~K.}\ \bibnamefont {{Jung}}}, \bibinfo {author}
  {\bibfnamefont {M.~J.}\ \bibnamefont {{Wilking}}}, \bibinfo {author}
  {\bibfnamefont {C.}~\bibnamefont {{Yanagisawa}}}, \bibinfo {author}
  {\bibfnamefont {M.}~\bibnamefont {{Harada}}}, \bibinfo {author}
  {\bibfnamefont {H.}~\bibnamefont {{Ishino}}}, \bibinfo {author}
  {\bibfnamefont {S.}~\bibnamefont {{Ito}}}, \bibinfo {author} {\bibfnamefont
  {H.}~\bibnamefont {{Kitagawa}}}, \bibinfo {author} {\bibfnamefont
  {Y.}~\bibnamefont {{Koshio}}}, \bibinfo {author} {\bibfnamefont
  {F.}~\bibnamefont {{Nakanishi}}}, \bibinfo {author} {\bibfnamefont
  {S.}~\bibnamefont {{Sakai}}}, \bibinfo {author} {\bibfnamefont
  {G.}~\bibnamefont {{Barr}}}, \bibinfo {author} {\bibfnamefont
  {D.}~\bibnamefont {{Barrow}}}, \bibinfo {author} {\bibfnamefont
  {L.}~\bibnamefont {{Cook}}}, \bibinfo {author} {\bibfnamefont
  {S.}~\bibnamefont {{Samani}}}, \bibinfo {author} {\bibfnamefont
  {D.}~\bibnamefont {{Wark}}}, \bibinfo {author} {\bibfnamefont
  {F.}~\bibnamefont {{Nova}}}, \bibinfo {author} {\bibfnamefont {J.~Y.}\
  \bibnamefont {{Yang}}}, \bibinfo {author} {\bibfnamefont {M.}~\bibnamefont
  {{Malek}}}, \bibinfo {author} {\bibfnamefont {J.~M.}\ \bibnamefont
  {{McElwee}}}, \bibinfo {author} {\bibfnamefont {O.}~\bibnamefont {{Stone}}},
  \bibinfo {author} {\bibfnamefont {M.~D.}\ \bibnamefont {{Thiesse}}}, \bibinfo
  {author} {\bibfnamefont {L.~F.}\ \bibnamefont {{Thompson}}}, \bibinfo
  {author} {\bibfnamefont {H.}~\bibnamefont {{Okazawa}}}, \bibinfo {author}
  {\bibfnamefont {S.~B.}\ \bibnamefont {{Kim}}}, \bibinfo {author}
  {\bibfnamefont {J.~W.}\ \bibnamefont {{Seo}}}, \bibinfo {author}
  {\bibfnamefont {I.}~\bibnamefont {{Yu}}}, \bibinfo {author} {\bibfnamefont
  {A.~K.}\ \bibnamefont {{Ichikawa}}}, \bibinfo {author} {\bibfnamefont
  {K.~D.}\ \bibnamefont {{Nakamura}}}, \bibinfo {author} {\bibfnamefont
  {S.}~\bibnamefont {{Tairafune}}}, \bibinfo {author} {\bibfnamefont
  {K.}~\bibnamefont {{Nishijima}}}, \bibinfo {author} {\bibfnamefont
  {K.}~\bibnamefont {{Iwamoto}}}, \bibinfo {author} {\bibfnamefont
  {K.}~\bibnamefont {{Nakagiri}}}, \bibinfo {author} {\bibfnamefont
  {Y.}~\bibnamefont {{Nakajima}}}, \bibinfo {author} {\bibfnamefont
  {N.}~\bibnamefont {{Taniuchi}}}, \bibinfo {author} {\bibfnamefont
  {M.}~\bibnamefont {{Yokoyama}}}, \bibinfo {author} {\bibfnamefont
  {K.}~\bibnamefont {{Martens}}}, \bibinfo {author} {\bibfnamefont
  {P.}~\bibnamefont {{de Perio}}}, \bibinfo {author} {\bibfnamefont {M.~R.}\
  \bibnamefont {{Vagins}}}, \bibinfo {author} {\bibfnamefont {M.}~\bibnamefont
  {{Kuze}}}, \bibinfo {author} {\bibfnamefont {S.}~\bibnamefont {{Izumiyama}}},
  \bibinfo {author} {\bibfnamefont {M.}~\bibnamefont {{Inomoto}}}, \bibinfo
  {author} {\bibfnamefont {M.}~\bibnamefont {{Ishitsuka}}}, \bibinfo {author}
  {\bibfnamefont {H.}~\bibnamefont {{Ito}}}, \bibinfo {author} {\bibfnamefont
  {T.}~\bibnamefont {{Kinoshita}}}, \bibinfo {author} {\bibfnamefont
  {R.}~\bibnamefont {{Matsumoto}}}, \bibinfo {author} {\bibfnamefont
  {Y.}~\bibnamefont {{Ommura}}}, \bibinfo {author} {\bibfnamefont
  {N.}~\bibnamefont {{Shigeta}}}, \bibinfo {author} {\bibfnamefont
  {M.}~\bibnamefont {{Shinoki}}}, \bibinfo {author} {\bibfnamefont
  {T.}~\bibnamefont {{Suganuma}}}, \bibinfo {author} {\bibfnamefont
  {K.}~\bibnamefont {{Yamauchi}}}, \bibinfo {author} {\bibfnamefont {J.~F.}\
  \bibnamefont {{Martin}}}, \bibinfo {author} {\bibfnamefont {H.~A.}\
  \bibnamefont {{Tanaka}}}, \bibinfo {author} {\bibfnamefont {T.}~\bibnamefont
  {{Towstego}}}, \bibinfo {author} {\bibfnamefont {R.}~\bibnamefont
  {{Akutsu}}}, \bibinfo {author} {\bibfnamefont {V.}~\bibnamefont
  {{Gousy-Leblanc}}}, \bibinfo {author} {\bibfnamefont {M.}~\bibnamefont
  {{Hartz}}}, \bibinfo {author} {\bibfnamefont {A.}~\bibnamefont {{Konaka}}},
  \bibinfo {author} {\bibfnamefont {N.~W.}\ \bibnamefont {{Prouse}}}, \bibinfo
  {author} {\bibfnamefont {S.}~\bibnamefont {{Chen}}}, \bibinfo {author}
  {\bibfnamefont {B.~D.}\ \bibnamefont {{Xu}}}, \bibinfo {author}
  {\bibfnamefont {B.}~\bibnamefont {{Zhang}}}, \bibinfo {author} {\bibfnamefont
  {M.}~\bibnamefont {{Posiadala-Zezula}}}, \bibinfo {author} {\bibfnamefont
  {D.}~\bibnamefont {{Hadley}}}, \bibinfo {author} {\bibfnamefont
  {M.}~\bibnamefont {{Nicholson}}}, \bibinfo {author} {\bibfnamefont
  {M.}~\bibnamefont {{O'Flaherty}}}, \bibinfo {author} {\bibfnamefont
  {B.}~\bibnamefont {{Richards}}}, \bibinfo {author} {\bibfnamefont
  {A.}~\bibnamefont {{Ali}}}, \bibinfo {author} {\bibfnamefont
  {B.}~\bibnamefont {{Jamieson}}}, \bibinfo {author} {\bibfnamefont
  {L.}~\bibnamefont {{Marti}}}, \bibinfo {author} {\bibfnamefont
  {A.}~\bibnamefont {{Minamino}}}, \bibinfo {author} {\bibfnamefont
  {G.}~\bibnamefont {{Pintaudi}}}, \bibinfo {author} {\bibfnamefont
  {S.}~\bibnamefont {{Sano}}}, \bibinfo {author} {\bibfnamefont
  {S.}~\bibnamefont {{Suzuki}}}, \bibinfo {author} {\bibfnamefont
  {K.}~\bibnamefont {{Wada}}},\ and\ \bibinfo {author} {\bibnamefont
  {{Super-Kamiokande Collaboration}}},\ }\bibfield  {title} {\bibinfo {title}
  {{Search for Cosmic-Ray Boosted Sub-GeV Dark Matter Using Recoil Protons at
  Super-Kamiokande}},\ }\href {https://doi.org/10.1103/PhysRevLett.130.031802}
  {\bibfield  {journal} {\bibinfo  {journal} {\prl}\ }\textbf {\bibinfo
  {volume} {130}},\ \bibinfo {eid} {031802} (\bibinfo {year} {2023})},\ \Eprint
  {https://arxiv.org/abs/2209.14968} {arXiv:2209.14968 [hep-ex]} \BibitemShut
  {NoStop}%
\bibitem [{\citenamefont {{Isern}}\ \emph {et~al.}(1992)\citenamefont
  {{Isern}}, \citenamefont {{Hernanz}},\ and\ \citenamefont
  {{Garcia-Berro}}}]{Isern1992}%
  \BibitemOpen
  \bibfield  {author} {\bibinfo {author} {\bibfnamefont {J.}~\bibnamefont
  {{Isern}}}, \bibinfo {author} {\bibfnamefont {M.}~\bibnamefont {{Hernanz}}},\
  and\ \bibinfo {author} {\bibfnamefont {E.}~\bibnamefont {{Garcia-Berro}}},\
  }\bibfield  {title} {\bibinfo {title} {{Axion Cooling of White Dwarfs}},\
  }\href {https://doi.org/10.1086/186416} {\bibfield  {journal} {\bibinfo
  {journal} {\apjl}\ }\textbf {\bibinfo {volume} {392}},\ \bibinfo {pages}
  {L23} (\bibinfo {year} {1992})}\BibitemShut {NoStop}%
\bibitem [{\citenamefont {{C{\'o}rsico}}\ \emph {et~al.}(2001)\citenamefont
  {{C{\'o}rsico}}, \citenamefont {{Benvenuto}}, \citenamefont {{Althaus}},
  \citenamefont {{Isern}},\ and\ \citenamefont
  {{Garc{\'\i}a-Berro}}}]{Corsico2001}%
  \BibitemOpen
  \bibfield  {author} {\bibinfo {author} {\bibfnamefont {A.~H.}\ \bibnamefont
  {{C{\'o}rsico}}}, \bibinfo {author} {\bibfnamefont {O.~G.}\ \bibnamefont
  {{Benvenuto}}}, \bibinfo {author} {\bibfnamefont {L.~G.}\ \bibnamefont
  {{Althaus}}}, \bibinfo {author} {\bibfnamefont {J.}~\bibnamefont {{Isern}}},\
  and\ \bibinfo {author} {\bibfnamefont {E.}~\bibnamefont
  {{Garc{\'\i}a-Berro}}},\ }\bibfield  {title} {\bibinfo {title} {{The
  potential of the variable DA white dwarf G117-B15A as a tool for fundamental
  physics}},\ }\href {https://doi.org/10.1016/S1384-1076(01)00055-0} {\bibfield
   {journal} {\bibinfo  {journal} {\na}\ }\textbf {\bibinfo {volume} {6}},\
  \bibinfo {pages} {197} (\bibinfo {year} {2001})},\ \Eprint
  {https://arxiv.org/abs/astro-ph/0104103} {arXiv:astro-ph/0104103 [astro-ph]}
  \BibitemShut {NoStop}%
\bibitem [{\citenamefont {{Niu}}\ and\ \citenamefont {{Xue}}(2022)}]{Niu2022}%
  \BibitemOpen
  \bibfield  {author} {\bibinfo {author} {\bibfnamefont {J.-S.}\ \bibnamefont
  {{Niu}}}\ and\ \bibinfo {author} {\bibfnamefont {H.-F.}\ \bibnamefont
  {{Xue}}},\ }\bibfield  {title} {\bibinfo {title} {{A Rapidly Evolving
  High-amplitude {\ensuremath{\delta}} Scuti Star Crossing the Hertzsprung
  Gap}},\ }\href {https://doi.org/10.3847/2041-8213/ac9407} {\bibfield
  {journal} {\bibinfo  {journal} {\apjl}\ }\textbf {\bibinfo {volume} {938}},\
  \bibinfo {eid} {L20} (\bibinfo {year} {2022})},\ \Eprint
  {https://arxiv.org/abs/2102.10259} {arXiv:2102.10259 [astro-ph.SR]}
  \BibitemShut {NoStop}%
\bibitem [{\citenamefont {{Xue}}\ \emph {et~al.}(2022)\citenamefont {{Xue}},
  \citenamefont {{Niu}},\ and\ \citenamefont {{Fu}}}]{Xue2022}%
  \BibitemOpen
  \bibfield  {author} {\bibinfo {author} {\bibfnamefont {H.-F.}\ \bibnamefont
  {{Xue}}}, \bibinfo {author} {\bibfnamefont {J.-S.}\ \bibnamefont {{Niu}}},\
  and\ \bibinfo {author} {\bibfnamefont {J.-N.}\ \bibnamefont {{Fu}}},\
  }\bibfield  {title} {\bibinfo {title} {{Precise Evolutionary Asteroseismology
  of High-Amplitude {\ensuremath{\delta}} Scuti Star AE Ursae Majoris}},\
  }\href {https://doi.org/10.1088/1674-4527/ac8b5e} {\bibfield  {journal}
  {\bibinfo  {journal} {Research in Astronomy and Astrophysics}\ }\textbf
  {\bibinfo {volume} {22}},\ \bibinfo {eid} {105006} (\bibinfo {year}
  {2022})},\ \Eprint {https://arxiv.org/abs/2208.09158} {arXiv:2208.09158
  [astro-ph.SR]} \BibitemShut {NoStop}%
\bibitem [{\citenamefont {{Bell}}\ \emph {et~al.}(2021)\citenamefont {{Bell}},
  \citenamefont {{Busoni}}, \citenamefont {{Ramirez-Quezada}}, \citenamefont
  {{Robles}},\ and\ \citenamefont {{Virgato}}}]{Bell2021}%
  \BibitemOpen
  \bibfield  {author} {\bibinfo {author} {\bibfnamefont {N.~F.}\ \bibnamefont
  {{Bell}}}, \bibinfo {author} {\bibfnamefont {G.}~\bibnamefont {{Busoni}}},
  \bibinfo {author} {\bibfnamefont {M.~E.}\ \bibnamefont {{Ramirez-Quezada}}},
  \bibinfo {author} {\bibfnamefont {S.}~\bibnamefont {{Robles}}},\ and\
  \bibinfo {author} {\bibfnamefont {M.}~\bibnamefont {{Virgato}}},\ }\bibfield
  {title} {\bibinfo {title} {{Improved treatment of dark matter capture in
  white dwarfs}},\ }\href {https://doi.org/10.1088/1475-7516/2021/10/083}
  {\bibfield  {journal} {\bibinfo  {journal} {\jcap}\ }\textbf {\bibinfo
  {volume} {2021}},\ \bibinfo {eid} {083} (\bibinfo {year} {2021})},\ \Eprint
  {https://arxiv.org/abs/2104.14367} {arXiv:2104.14367 [hep-ph]} \BibitemShut
  {NoStop}%
\bibitem [{\citenamefont {{Garani}}\ \emph {et~al.}(2019)\citenamefont
  {{Garani}}, \citenamefont {{Genolini}},\ and\ \citenamefont
  {{Hambye}}}]{Garani2019}%
  \BibitemOpen
  \bibfield  {author} {\bibinfo {author} {\bibfnamefont {R.}~\bibnamefont
  {{Garani}}}, \bibinfo {author} {\bibfnamefont {Y.}~\bibnamefont
  {{Genolini}}},\ and\ \bibinfo {author} {\bibfnamefont {T.}~\bibnamefont
  {{Hambye}}},\ }\bibfield  {title} {\bibinfo {title} {{New analysis of neutron
  star constraints on asymmetric dark matter}},\ }\href
  {https://doi.org/10.1088/1475-7516/2019/05/035} {\bibfield  {journal}
  {\bibinfo  {journal} {\jcap}\ }\textbf {\bibinfo {volume} {2019}},\ \bibinfo
  {eid} {035} (\bibinfo {year} {2019})},\ \Eprint
  {https://arxiv.org/abs/1812.08773} {arXiv:1812.08773 [hep-ph]} \BibitemShut
  {NoStop}%
\bibitem [{\citenamefont {{Busoni}}\ \emph {et~al.}(2017)\citenamefont
  {{Busoni}}, \citenamefont {{De Simone}}, \citenamefont {{Scott}},\ and\
  \citenamefont {{Vincent}}}]{Busoni2017}%
  \BibitemOpen
  \bibfield  {author} {\bibinfo {author} {\bibfnamefont {G.}~\bibnamefont
  {{Busoni}}}, \bibinfo {author} {\bibfnamefont {A.}~\bibnamefont {{De
  Simone}}}, \bibinfo {author} {\bibfnamefont {P.}~\bibnamefont {{Scott}}},\
  and\ \bibinfo {author} {\bibfnamefont {A.~C.}\ \bibnamefont {{Vincent}}},\
  }\bibfield  {title} {\bibinfo {title} {{Evaporation and scattering of
  momentum- and velocity-dependent dark matter in the Sun}},\ }\href
  {https://doi.org/10.1088/1475-7516/2017/10/037} {\bibfield  {journal}
  {\bibinfo  {journal} {\jcap}\ }\textbf {\bibinfo {volume} {2017}},\ \bibinfo
  {eid} {037} (\bibinfo {year} {2017})},\ \Eprint
  {https://arxiv.org/abs/1703.07784} {arXiv:1703.07784 [hep-ph]} \BibitemShut
  {NoStop}%
\bibitem [{\citenamefont {{Gould}}(1987)}]{Gould1987b}%
  \BibitemOpen
  \bibfield  {author} {\bibinfo {author} {\bibfnamefont {A.}~\bibnamefont
  {{Gould}}},\ }\bibfield  {title} {\bibinfo {title} {{Weakly interacting
  massive particle distribution in and evaporation from the sun}},\ }\href
  {https://doi.org/10.1086/165652} {\bibfield  {journal} {\bibinfo  {journal}
  {\apj}\ }\textbf {\bibinfo {volume} {321}},\ \bibinfo {pages} {560} (\bibinfo
  {year} {1987})}\BibitemShut {NoStop}%
\bibitem [{\citenamefont {{Garani}}\ and\ \citenamefont
  {{Palomares-Ruiz}}(2017)}]{Garani2017}%
  \BibitemOpen
  \bibfield  {author} {\bibinfo {author} {\bibfnamefont {R.}~\bibnamefont
  {{Garani}}}\ and\ \bibinfo {author} {\bibfnamefont {S.}~\bibnamefont
  {{Palomares-Ruiz}}},\ }\bibfield  {title} {\bibinfo {title} {{Dark matter in
  the Sun: scattering off electrons vs nucleons}},\ }\href
  {https://doi.org/10.1088/1475-7516/2017/05/007} {\bibfield  {journal}
  {\bibinfo  {journal} {\jcap}\ }\textbf {\bibinfo {volume} {2017}},\ \bibinfo
  {eid} {007} (\bibinfo {year} {2017})},\ \Eprint
  {https://arxiv.org/abs/1702.02768} {arXiv:1702.02768 [hep-ph]} \BibitemShut
  {NoStop}%
\end{thebibliography}
%apsrev4-2.bst 2019-01-14 (MD) hand-edited version of apsrev4-1.bst
%Control: key (0)
%Control: author (8) initials jnrlst
%Control: editor formatted (1) identically to author
%Control: production of article title (0) allowed
%Control: page (0) single
%Control: year (1) truncated
%Control: production of eprint (0) enabled
%

\clearpage
\appendix
\section{Capture and Evaporation rates of DM Particles}
\label{app:01}

In general, there is an absolute upper limit on the capture rate when a WD capture DM particles, in which case every DM particle traversing the star is captured. This saturation capture condition is called the geometric limit of the capture rate, which is given by \cite{Bell2021}
\begin{equation}
\label{eq:cap_geom}
C_\mathrm{geo}= \frac{\pi \Rstar^{2}}{3\vstar} \frac{\rhodm}{\mdm} \left[ (3 \vesc^{2}(\Rstar)+3\vstar^{2}+v_{d}^{2}) \cdot \mathrm{Erf} \left(\sqrt{\frac{3}{2}} \frac{\vstar}{v_{d}} \right) + \sqrt{\frac{6}{\pi}} \vstar v_{d} \cdot \exp \left( - \frac{3\vstar^{2}}{2v_{d}^{2}} \right) \right],
\end{equation}
where $\rhodm = 0.3\ \GeV/\cm^{3}$ is the local DM density around the Sun; $\mdm$ is mass of the DM particle; $\Rstar$ is the radius of the star; $\vesc (r)$ is the escape velocity of a DM particle from a distance $r$ to the center of the star; $\vstar = 220\ \km/\s$ is the relative velocity of the Sun to the frame where DM has null average speed; $v_{d} = 270 \ \km/\s$ is the dispersion velocity of DM particles near the Sun. Because the distances from the Sun to the three DAVs can be ignored compared to the galaxy center ($\sim 8.5 \kpc$), we use the above values of $\rhodm$, $\vstar$, and $v_{d}$ all through this work.

In the non-saturation capture condition, the capture rate can be written as \cite{Garani2019,Bell2021}
\begin{equation}
  \label{eq:cap_non}
  C_\mathrm{ngeo}= \int_{0}^{\Rstar} \frac{\rhodm}{\mdm} 4\pi r^{2} \diff r \int_{0}^{\infty} \frac{f_{\vstar}(\udm)}{\udm} w(r) \diff \udm \int_{0}^{\vesc} R^{-}(w \rightarrow v) \diff v,
\end{equation}
where $\udm$ is the velocity of DM particles far from the star; $w(r) = \sqrt{\udm^{2} + \vesc^{2}(r)}$ is the velocity of a DM particle at a finite distance from the center of the star; $R^{-}(w \rightarrow v)$ is the differential scattering rate for capture (in the star's frame for a DM particle with velocity $w$ to scatter to a smaller velocity $v\ (w>v)$), which depends on the equation of state of the constituent in the star and is different for nuclei and electrons; $f_{\vstar}(\udm)$ is the DM velocity distribution, which is usually assumed to be a Maxwell-Boltzmann distribution and expressed as \cite{Busoni2017}
\begin{equation}
\label{eq:fDM_out}
f_{\vstar}(\udm) = \frac{\udm}{\vstar} \sqrt{\frac{3}{2\pi (v_{d}^{2} + 3kT_{*}/m_{T})}} \left( \exp \left[- \frac{3(\udm - \vstar)^{2}}{2(v_{d}^{2} + 3kT_{*}/m_{T})} \right] -  \exp \left[- \frac{3(\udm + \vstar)^{2}}{2(v_{d}^{2} + 3kT_{*}/m_{T})} \right] \right),
\end{equation}
where $k$ is the Boltzmann constant; $T_{*}$ is the temperature of the star, which depend on $r$ in general; $m_{T}$ is the mass of the target particles (nuclei or electrons).

The capture rate in general is defined as 
\begin{equation}
  \label{eq:cap}
  C_\mathrm{*}= \min \{C_\mathrm{geo}, C_\mathrm{ngeo}\}.
\end{equation}

The evaporation rate can be written as \cite{Gould1987b,Garani2017,Busoni2017}
\begin{equation}
  \label{eq:eva}
  E_{*}= \int_{0}^{\Rstar} \ndm (r) 4\pi r^{2} \diff r \int_{0}^{\vesc} f_{\chi}(w,r) 4\pi w^{2} \diff w \int_{\vesc}^{\infty} R^{+}(w \rightarrow v) \diff v,
\end{equation}
where $R^{+}(w \rightarrow v)$ is the differential scattering rate for evaporation, which depends on the target particles in the star; $\ndm (r)$ is the normalized radial distribution of DM \cite{Gould1987b,Garani2017,Garani2019}:
\begin{equation}
  \label{eq:DM_dis}
  \ndm (r) = \frac{4}{\rdm^{3} \sqrt{\pi}} \exp \left(- \frac{r^2}{\rdm^{2}} \right),\text{with}\  \rdm = \sqrt{\frac{3kT_{*}}{2 \pi G \rhostar \mdm}},
\end{equation}
where $G$ is the gravitational constant, and $\rhostar$ is the density of the star; $f_{\chi}(w,r)$ is the thermalized DM particles' velocity distribution, follows a Maxwell-Boltzmann distribution truncated at the escape velocity $\vesc(r)$ given by \cite{Gould1987b,Garani2017,Garani2019}
\begin{equation}
\label{eq:fDM_out}
f_{\chi}(w,r) = \frac{1}{\pi^{3/2}} \left(\frac{\mdm}{2kT_{*}}\right)^{3/2} \frac{\exp \left(-\frac{\mdm w^{2}}{2kT_{*}} \right) \Theta(\vesc(r) - w)}{\mathrm{Erf} \left(\sqrt{\frac{\mdm \vesc^{2}(r)}{2kT_{*}}} \right) - \frac{2}{\sqrt{\pi}} \sqrt{\frac{\mdm \vesc^{2}(r)}{2kT_{*}}} \exp \left(-\frac{\mdm \vesc^{2}(r)}{2kT_{*}} \right)}.
\end{equation}

For nuclei, the differential scattering rates for capture and evaporation can be expressed as \cite{Busoni2017}
\begin{equation}
  \label{eq:R_cap_n}
  R^{-}(w \rightarrow v)= \frac{32\mu_{+}^{4}}{\sqrt{\pi}} \kappa^{3} n_{n}(r) \frac{\diff \sigma_{N}}{\diff \cos \theta} \frac{v}{w} \int_{0}^{\infty} \diff \vs \int_{0}^{\infty}  \vt e^{-\kappa^{2} v_{T}^{2}} H^{-}(\vs,\vt,w,v) \diff \vt,
\end{equation}
and
\begin{equation}
  \label{eq:R_eva_n}
  R^{+}(w \rightarrow v)= \frac{32\mu_{+}^{4}}{\sqrt{\pi}} \kappa^{3} n_{n}(r) \frac{\diff \sigma_{N}}{\diff \cos \theta} \frac{v}{w} \int_{0}^{\infty} \diff \vs \int_{0}^{\infty}  \vt e^{-\kappa^{2} v_{T}^{2}} H^{+}(\vs,\vt,w,v) \diff \vt,
\end{equation}
with
\begin{align}
  \label{eq:notion_n}
  \begin{aligned}
  \mu_{\pm} &= \frac{\mu \pm 1}{2}, \mu = \frac{\mdm}{m_{n}}, \kappa^{2} = \frac{m_{n}}{2kT_{*}},\\
  v_{T}^{2} &= 2\mu \mu_{+} \vt^{2} + 2\mu \vs^{2} - \mu w^{2}, \\
  H^{-}(\vs,\vt,w,v) &= \Theta(\vt+\vs-w) \Theta(v-|\vt-\vs|), \\
  H^{+}(\vs,\vt,w,v) &= \Theta(\vt+\vs-v) \Theta(w-|\vt-\vs|),
  \end{aligned}
\end{align}
where $n_{n}(r)$ is the number density of the nucleus; $\vs$ is the velocity in the star's frame of the center of mass (CM) of the scattering, $\vt$ is the velocity of DM in the CM frame; $m_{n}$ is the mass of the nucleus. 
In this work, we choose the constant differential cross section form:
\begin{equation}
  \label{eq:cs_diff}
  \frac{\diff \sigma_{N}}{\diff \cos \theta} = \frac{\sigma_{N}}{2}, 
\end{equation}
with
\begin{equation}
  \label{eq:cs_n}
  \sigma_{N}=\sign A^{2} \left(\frac{m_{n}}{m_{p}} \right)^{2} \left(\frac{\mdm + m_{p}}{\mdm + m_{n}} \right)^{2},
\end{equation}
where $\sign$ is the spin-independent cross section between DM particle and nucleon, which is assumed to be the same for proton and neutron; $A$ is the mass number of the nucleus.

For electrons, the differential scattering rates for capture and evaporation can be expressed as \cite{Garani2019}
\begin{equation}
  \label{eq:R_cap_e}
  R^{-}(w \rightarrow v)= 8 \mu_{+}^{4} \sige F(q) n_{e}(r) \frac{v}{w} \int_{0}^{\infty} \diff \vs \int_{0}^{\infty}  \vt f_{p}(E_{p},r) (1-f_{p'}(E_{p'},r)) H^{-}(\vs,\vt,w,v) \diff \vt,
\end{equation}
and 
\begin{equation}
  \label{eq:R_eva_e}
  R^{+}(w \rightarrow v)= 8 \mu_{+}^{4} \sige F(q) n_{e}(r) \frac{v}{w} \int_{0}^{\infty} \diff \vs \int_{0}^{\infty}  \vt f_{p}(E_{p},r) (1-f_{p'}(E_{p'},r)) H^{+}(\vs,\vt,w,v) \diff \vt,
\end{equation}
with
\begin{align}
  \label{eq:notion_e}
  \begin{aligned}
    \mu_{\pm} &= \frac{\mu \pm 1}{2}, \mu = \frac{\mdm}{m_{e}}\\
    f_{p}(E_{p},r) &= \left( \exp \left(\frac{E_{p}-\mu_{F}(r)}{T_{*}} \right) +1 \right)^{-1},\\
    1-f_{p'}(E_{p'},r) &= 1- \left( \exp \left(\frac{E_{p'}-\mu_{F}(r)}{T_{*}} \right) +1 \right)^{-1},\\
    E_{p} &= \frac{1}{2} m_{e} (2\mu \mu_{+} \vt^{2} + 2\mu \vs^{2} - \mu w^{2}),\\
    E_{p'} &= \frac{1}{2} m_{e} (2\mu \mu_{+} \vt^{2} + 2\mu \vs^{2} - \mu v^{2}),
  \end{aligned}
\end{align}
where $n_{e}(r)$ is the number density of the electrons; $m_{e}$ is the mass of the electron; $f_{p}(E_{p},r)$ and $1-f_{p'}(E_{p'},r)$ is the Fermi-Dirac distribution for the electron in the initial and final state; $\mu_{F}(r)$ is the chemical potential of the electron at $r$; $H^{\pm}(\vs,\vt,w,v)$ is the same expression as that in Eq. (\ref{eq:notion_n}); $F(q)$ is the momentum-dependent DM form factor. 
In this work, we consider two cases: $F(q) = 1$ and $F(q) = (\alpha m_{e})^{2}/q^{2}$, where $\alpha$ is the  fine-structure constant and $q$ is the transferred momentum.

Besides, some additional simplifications are performed as well:  (i) a uniform distribution of matters in a DAV: $\rhostar (r) = \rhostar = \Mstar/\Vstar$ ($\rhostar$, $\Mstar$ and $\Vstar$ are the density, mass and volume of a DAV); (ii) the same chemical composition over the entire scattering volume $\Vstar$ (because all the three DAVs in this work are carbon-oxygen core WDs, we use a mean atomic weight 14 for the nucleus \cite{Mukadam2013}); (iii) a uniform temperature profile ($T_{*}$ does not depend on $r$) in a DAV because of the extremely high thermal conductivity of an electron degenerate core; (iv) as DAVs are always electrically neutral, we use the value ${\Mstar}/{2m_{p}}$ and ${\rhostar}/2{m_{p}}$ as the total number of electrons ($N_{e}$) and number density of electrons ($n_{e}$) in a DAV, respectively ($m_{p}$ is the mass of proton). 

\section{Energy Transferred by DM Capture and Evaporation}
\label{app:02}
The energy transferred by the capture process (DM particles transfer kinetic energy to the nuclei/electrons in the star) and evaporation process (the nuclei/electrons in the star transfer kinetic energy to the DM particles) in a unit of time can be written as
\begin{align}
  \label{eq:Ein}
  \begin{aligned}
    \Ein &= \frac{C_\mathrm{geo}}{C_\mathrm{ngeo}}\int_{0}^{\Rstar} \frac{\rhodm}{\mdm} 4\pi r^{2} \diff r \int_{0}^{\infty} \frac{f_{\vstar}(\udm)}{\udm} w(r) \diff \udm \int_{0}^{\vesc} \frac{1}{2} \mdm (w^2 - v^2) \cdot R^{-}(w \rightarrow v) \diff v, \\
    &\text{for} \quad  C_\mathrm{geo} < C_\mathrm{ngeo}; \\
    \Ein &= \int_{0}^{\Rstar} \frac{\rhodm}{\mdm} 4\pi r^{2} \diff r \int_{0}^{\infty} \frac{f_{\vstar}(\udm)}{\udm} w(r) \diff \udm \int_{0}^{\vesc} \frac{1}{2} \mdm (w^2 - v^2) \cdot R^{-}(w \rightarrow v) \diff v, \\
    &\text{for} \quad  C_\mathrm{geo} \ge C_\mathrm{ngeo};
  \end{aligned}
\end{align}
and
\begin{equation}
  \label{eq:Eout}
  \Eout = C_{*} \int_{0}^{\Rstar} \ndm (r) 4\pi r^{2} \diff r \int_{0}^{\vesc} f_{\chi}(w,r) 4\pi w^{2} \diff w \int_{\vesc}^{\infty} \frac{1}{2} \mdm (v^2 - w^2) \cdot R^{+}(w \rightarrow v) \diff v,
\end{equation}
which can be obtained based on Eqs. (\ref{eq:R_cap_n}), (\ref{eq:R_eva_n}), (\ref{eq:R_cap_e}), and (\ref{eq:R_eva_e}).

\end{document}